\documentclass[sigconf,nonacm]{acmart}
\setcopyright{none}
\renewcommand\footnotetextcopyrightpermission[1]{}
\AtBeginDocument{%
  }

\def\eqref#1{equation~\ref{#1}}

\def\1{\bm{1}}

\def\sA{{\mathbb{A}}}

\def\sD{{\mathbb{D}}}
\def\sE{{\mathbb{E}}}
\def\sF{{\mathbb{F}}}

\def\sN{{\mathbb{N}}}

\def\sP{{\mathbb{P}}}
\def\sQ{{\mathbb{Q}}}
\def\sR{{\mathbb{R}}}
\def\sS{{\mathbb{S}}}
\def\sT{{\mathbb{T}}}

\def\sW{{\mathbb{W}}}

\usepackage{multirow}
\usepackage{placeins} 
\usepackage{array}
\usepackage{pifont}
\usepackage{enumitem}

\newcommand{\cmark}{\ding{51}} 
\newcommand{\xmark}{\ding{55}} 

\begin{document}

\title{E-CARE: An Efficient LLM-based Commonsense-Augmented Framework for E-Commerce}

\author{Ge Zhang}
\email{ge.zhang1@huawei.com}
\affiliation{%
  \institution{Huawei Noah's Ark Lab}
  \city{Montreal}
  \state{Québec}
  \country{Canada}
}

\author{Rohan Deepak Ajwani}
\email{rohan.ajwani@gmail.com}
\affiliation{%
  \institution{Huawei Noah's Ark Lab}
  \city{Montreal}
  \state{Québec}
  \country{Canada}
}


\author{Yaochen Hu}
\email{yaochen.hu@huawei.com}
\affiliation{%
  \institution{Huawei Noah's Ark Lab}
  \city{Montreal}
  \state{Québec}
  \country{Canada}
}

\author{Tony Zheng}
\email{t28zheng@uwaterloo.ca}
\affiliation{%
  \institution{Huawei Noah's Ark Lab}
  \city{Montreal}
  \state{Québec}
  \country{Canada}
}

\author{Hongjian Gu}
\email{hongjian.gu@huawei.com}
\affiliation{%
  \institution{Huawei Noah's Ark Lab}
  \city{Montreal}
  \state{Québec}
  \country{Canada}
}

\author{Wei Guo}
\email{guowei67@huawei.com}
\affiliation{%
  \institution{Huawei Noah's Ark Lab}
  \country{Singapore}
}

\author{Mark Coates}
\email{mark.coates@mcgill.ca}
\affiliation{%
  \institution{McGill University}
  \institution{Mila - Québec AI Institute}
  \city{Montreal}
  \state{Québec}
  \country{Canada}
}

\author{Yingxue Zhang}
\email{yingxue.zhang@huawei.com}
\affiliation{%
  \institution{Huawei Noah's Ark Lab}
  \city{Montreal}
  \state{Québec}
  \country{Canada}
}

\renewcommand{\shortauthors}{Zhang et al.}

\begin{abstract}
Finding relevant products given a user query is pivotal to an e-commerce platform, as it can drive shopping behavior and generate revenue. The challenge lies in accurately predicting the correlation between queries and products. Recently, mining commonsense knowledge between queries and products using Large Language Models (LLMs) has shown promising results in boosting recommendation performance. However, such methods incur high costs due to intensive real-time LLM decoding during inference, as well as human annotation and potential Supervised Fine-Tuning (SFT) during training. To boost efficiency while leveraging LLMs' commonsense reasoning for various e-commerce tasks, we propose the Efficient Commonsense-Augmented Recommendation Enhancer (E-CARE), which requires neither SFT nor human annotation. The recommendation models augmented with E-CARE can access commonsense reasoning by leveraging a reasoning factor graph that encodes most of the reasoning schema from powerful LLMs, without requiring real-time LLM decoding. The experiments on 2 downstream tasks show improvements of up to 12.1\% in precision@5.
\end{abstract}






\maketitle

\section{Introduction}

Finding relevant products given a user query efficiently is pivotal for E‑commerce platforms \citep{kumar2021NeuralSearchLearning, shang2025KnowledgeDistillationEnhancing}. Early E‑commerce search systems relied on lexical matching methods such as TFIDF \citep{ramosUsingTFIDFDetermine}, and BM25 \citep{keenOkapiTREC3}. Although computationally efficient, these approaches match queries and product descriptions at the term level and cannot bridge the lexical gap that arises when users describe products with different words. Bi‑encoder \citep{reimers2019SentenceBERTSentenceEmbeddings} models (also called two‑tower or Siamese models) resolve the lexical gap by learning embeddings for queries and products separately and evaluating the relevance by the similarity of those embeddings. They offer high throughput thanks to offline computation of product embeddings and efficient approximate nearest neighbor search systems such as Faiss \citep{douze2025Faisslibrary}. To further improve performance, cross‑encoder models \citep{wu2022PracticeImprovingSearch} evaluate the relevance between the query and product by jointly feeding the query and product text into a trainable model, allowing the encoder to examine interactions across all tokens and potentially achieving better performance. However, cross-encoders are significantly slower than bi-encoders because they require running an encoder for each query-product pair, which poses challenges for real-time retrieval of millions of products \citep{schlatt2024InvestigatingEffectsSparse}. 

\begin{figure}[t]
    \centering
    \includegraphics[width=\columnwidth]{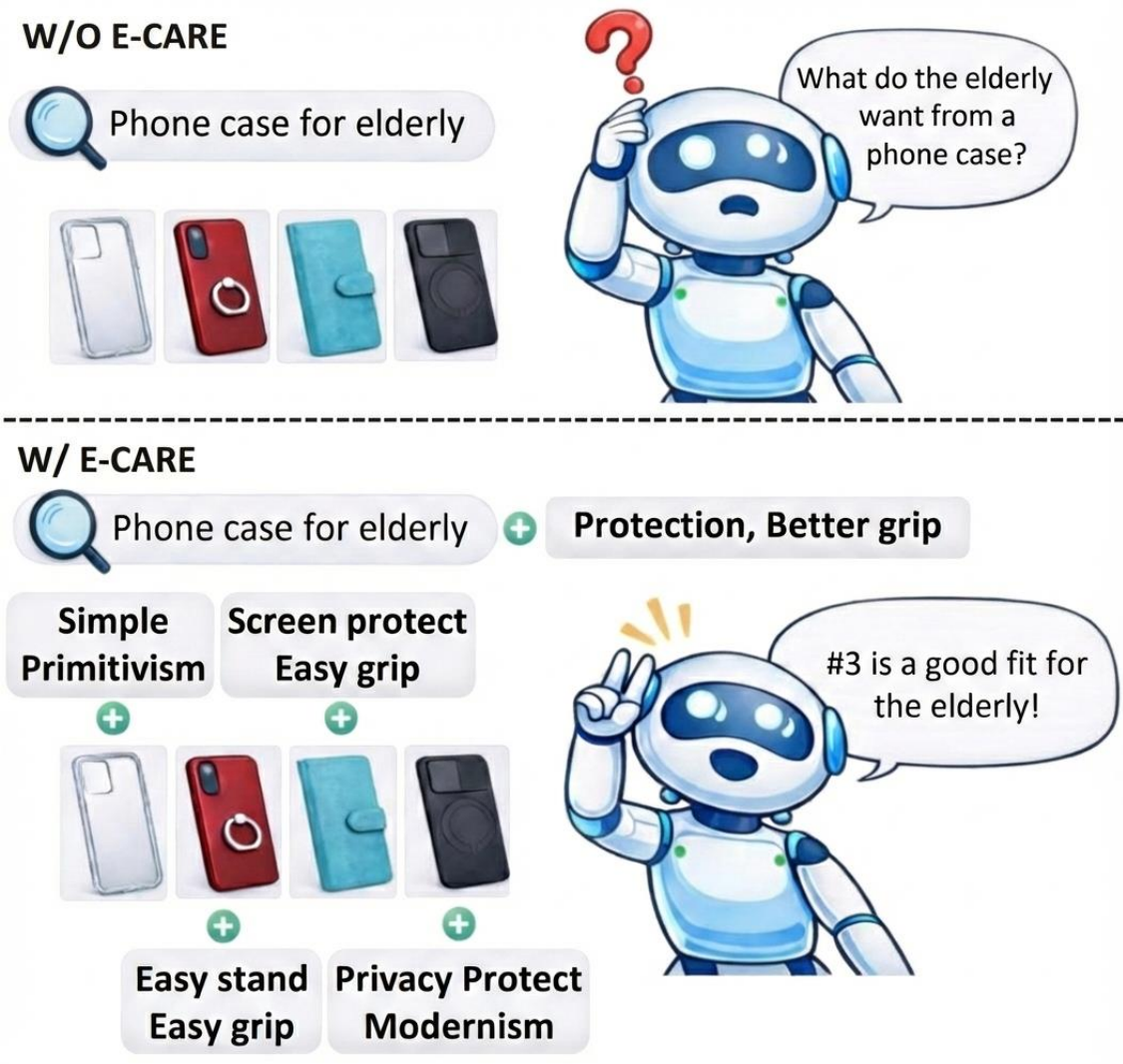}
    \caption{Recommendation systems struggle when queries provide vague requirements without specifying the features or functionalities that are needed. By augmenting both query and products with extra knowledge generated with commonsense perspectives from historical interactions by LLMs, recommendation systems can find correlations more easily.}
    \label{fig:overall_diagram}
\end{figure}

Despite the increasing capacity and complexity of models, these early attempts are insufficient in real-world scenarios. In practice, queries can be vague, failing to specify product features in detail, and semantic similarities alone may be insufficient to determine whether a product satisfies a query. For example, as shown in Figure \ref{fig:overall_diagram}, a query such as ``phone case for the elderly'' doesn't specify any properties, such as brand or functionality, that could inform recommendations. However, from a commonsense perspective, ``the elderly'' implies that the user may prioritize practical considerations, such as protection and grip, over aesthetics. Fulfilling such implicit expectations requires not only lexical and semantic features from the original query and product texts, but also commonsense reasoning based on cross-features ~\citep{li2023EcomGPTInstructiontuningLarge, yu2023FolkScopeIntentionKnowledgea, yu2024COSMOLargeScaleEcommercec}. 

Large Language Models (LLMs) have demonstrated their capabilities for commonsense reasoning across various tasks \citep{wei2023ChainofThoughtPromptingElicitsa, wang2024ChainThoughtReasoning, kojima2023LargeLanguageModels, zhao2023LargeLanguageModels, openai2024OpenAIo1System}, paving the way for integrating commonsense reasoning into query-product recommendation scenarios. Previous work like RepLLaMA \citep{ma2023FineTuningLLaMAMultiStagea} shows improvements in text retrieval scenarios by utilizing LLMs as encoders in a Bi-encoder framework, while other techniques like RankGPT \citep{sun2023ChatGPTGoodSearch} and RankVicuna \citep{pradeep2023RankVicunaZeroShotListwisea} directly rank items through a prompting approach.

The recently proposed methods FolkScope \citep{yu2023FolkScopeIntentionKnowledgea} and COSMO \citep{yu2024COSMOLargeScaleEcommercec} exploit the commonsense reasoning power of LLMs by jointly analyzing the query and product pairs and augmenting the main relevance prediction model with the reasoning results. Although they make effective use of LLMs, the methods are slow and costly. They rely heavily on human annotators and Supervised Fine-Tuning (SFT) during training and on real-time LLM decoding during inference, which is not scalable in high-throughput scenarios with a large number of candidate items.

In this paper, we propose Efficient Commonsense Augmented Recommendation Enhancer (E-CARE), a novel paradigm that enables efficient access to commonsense reasoning knowledge for queries and products comparable to invoking LLMs for each individual query-product pair. First, a commonsense reasoning factor graph is constructed from historical query-product interactions to encapsulate the underlying reasoning factors generated by powerful LLMs. Second, lightweight adapters are trained to learn correlations between queries/products and reasoning factors on the graph. Finally, during the inference stage, downstream recommendation systems can efficiently leverage LLM-level commonsense reasoning via adapters rather than costly LLMs. 

Furthermore, we introduce a 3-stage pipeline to generate a reasoning factor graph without relying on heavy components such as SFT or human annotation. Specifically, the pipeline comprises 1) \textit{Reasoning}, which leverages LLMs to mine commonsense reasoning factors to initialize the reasoning factor graph based on both query-product cross interactions and item inherent features; 2) \textit{Node Clustering}, which clusters and aggregate similar reasoning factors; and 3) \textit{Edge Filtering}, that removes less confident edges of the graph through LLM uncertainty evaluation to improve the quality and reduce the size of the graph.

Our contributions include:
\begin{itemize}
    \item A novel paradigm that utilizes the reasoning capacity of LLMs efficiently. E-CARE efficiently extracts commonsense reasoning knowledge and formalizes it as a graph. With lightweight adapters, downstream recommendation models can access LLM-level commonsense reasoning knowledge during inference without real-time LLM decoding. 
    \item A 3-stage LLM-based pipeline that can generate the reasoning factor graph without costly components, such as SFT and human annotation.
    \item Empirical experiments of E-CARE on 2 e‑commerce recommendation tasks, search relevance and app recall, demonstrate the improvements up to 12.79\% on Macro F1 and 12.1\% on Recall@5, respectively.
\end{itemize}

\section{Related Work} \label{sec2} 

In this section, we review the literature on existing retrieval and recommendation systems. We first briefly review conventional retrieval methods that use dense vector representations. Next, we discuss LLM-based methods that use zero- or few-shot instructions to output the items. Finally, we review methods that enhance retrieval and recommendation by LLMs' reasoning ability.

\subsection{Conventional Retrieval Methods}

Dense retrievers aim to recall a relevant subset of items from dense vector representations, typically generated by transformer-based language models such as BERT \citep{devlinBERTPretrainingDeep} and T5 \citep{raffelExploringLimitsTransfer}. Based on the model architecture, these dense retrievers can be broadly classified into two categories: bi-encoders and cross-encoders. Bi-encoders \citep{karpukhin2020DensePassageRetrievala, yu2021FewShotConversationalDense, xiong2021ANSWERINGCOMPLEXOPENDOMAIN, li2021MoreRobustDense, xiong2021APPROXIMATENEARESTNEIGHBOR, lin2021InBatchNegativesKnowledge, huang2013Learningdeepstructured, humeau2020Polyencodersarchitecturespretraining} use a two-tower structure that encodes the query and item text separately, enabling efficient retrieval over the entire candidate set. On the other hand, cross-encoder architectures \citep{nogueira2020PassageRerankingBERT, nogueira2020DocumentRankingPretrained, dai2019DeeperTextUnderstanding, zou2022DivideConquerText} jointly encode the concatenated query and item texts into a single embedding, which is then used for high-accuracy classification or re-ranking. Late interaction models such as ColBERT \citep{khattab2020ColBERTEfficientEffective} and its subsequent improvements \citep{santhanam2022ColBERTv2EffectiveEfficient, hofstatter2022IntroducingNeuralBag} act as a hybrid between bi-encoder and cross-encoder architectures. They retain token-level interactions between queries and items while maintaining efficiency by precomputing query and item embeddings offline. Several works, such as MADR~\citep{kong2022MultiAspectDenseRetrievala}, AGREE~\citep{shan2023TwoTowerAttributeGuidedb}, and SANTA~\citep{li2023StructureAwareLanguageModela}, extend beyond merely using query and item representations, and incorporate item features to produce recommendations.

While these conventional dense retrieval methods are faster and lighter than LLMs, they rely on the semantic similarity between queries and items. They are either unable to capture or are simply unaware of fine-grained nuances and relationships between texts. This restricts their ability to generalize, especially for niche items and long-tail queries.

\subsection{LLMs Directly as Classifier or Ranker}
\label{relatedwork_direct_llm}

Several works propose prompting pre-trained large language models \citep{brownLanguageModelsare, openai2024GPT4TechnicalReporta, touvron2023LLaMAOpenEfficienta} to classify or rerank items given a query. \citet{sachan2022ImprovingPassageRetrieval} uses a zero-shot instruction as a prompt to generate the probability of the query given the passage, and uses this probability to rank the passage. \citet{qin2024LargeLanguageModels} prompt LLMs to output more relevant passages to a query given a pair of passages, using three approaches for ranking, viz., all-pair comparisons, sorting-based, and sliding window. They swap the order of pairs in the prompt for each pair to de-bias the results. RankGPT \citep{sun2023ChatGPTGoodSearch} uses a zero-shot prompt, with a role-playing instruction component, ``You are RankGPT, an intelligent assistant that can rank passages [...]'', with GPT-4 \citep{openai2024GPT4TechnicalReporta} to generate a ranked list of passages given a query and an unordered list of passages. Other listwise ranking methods include LRL \citep{ma2023ZeroShotListwiseDocument} and RankVicuna \citep{pradeep2023RankVicunaZeroShotListwisea}. \citet{zhuang2024SetwiseApproachEffective} propose a setwise prompting approach that provides LLMs with a query and a set of items for ranking. This approach retains the effectiveness of pairwise ranking while significantly reducing the number of LLM calls, thereby enhancing efficiency. 

Despite these advances, existing prompting-based ranking methods still face significant latency issues due to inevitable real-time LLM decoding. Moreover, they do not explicitly capture user intent or item attribute relevance, making them less effective for queries that require commonsense reasoning. In addition, these methods generally lack interpretability, as they provide little insight into why one output is preferred over another. 

\subsection{LLMs as Reasoner}

While zero-shot prompt-based ranking (\S \ref{relatedwork_direct_llm}) has shown impressive performance, it still lags behind sophisticated supervised dense retrieval architectures \citep{ma2023ZeroShotListwiseDocument}. To overcome this, recent work proposes using LLMs' reasoning abilities to re-rank and classify relevant items. RaCT \citep{liu2025RaCTRankingawareChainofThought} utilizes CoT \citep{wei2023ChainofThoughtPromptingElicitsa} to guide the model to iteratively rank passages by relevance to a query. They incorporate this prompting into Ranking Preference Optimization \citep{rafailovDirectPreferenceOptimization}. Rank-R1 \citep{zhuang2025RankR1EnhancingReasoning} uses GRPO \citep{shao2024DeepSeekMathPushingLimits} to train LLMs to generate reasoning steps before selecting the most relevant item from a set of items (setwise ranking). \citet{qin2025TongSearchQRReinforcedQuery} propose TongSearch-QR, which trains small-scale language models using RL to rewrite a query with reasoning, which is then used to retrieve relevant items. RankCoT \citep{wu2025RankCoTRefiningKnowledge} trains language models to generate CoT-based summaries of retrieved items, which are then used to re-rank the items based on their relevance to the query. 
However, these reasoning-based methods either depend on nontrivial reinforcement learning procedures or incur substantial LLMs decoding overhead due to long CoT generations or iterative comparisons, which hurts scalability on large candidate sets.

Other works extend LLMs' reasoning abilities to generate intention graphs that map user queries to products, leveraging LLMs' commonsense reasoning to form query-product connections. Folk-Scope \citep{yu2023FolkScopeIntentionKnowledgea} utilizes LLMs to generate intention assertions using prompts for e-commerce, aiming to explain shopping behaviors. The generated knowledge is manually annotated and condensed into a structured knowledge graph, which is then used for downstream e-commerce applications. COSMO \citep{yu2024COSMOLargeScaleEcommercec} utilizes instruction tuning to finetune COSMO-LM, which then generates commonsense knowledge in e-commerce based on query-item relevance prompts. This fine-tuned language model is used to generate a knowledge graph, whose features are then utilized to enhance search results.

Reasoning-based methods, while accurate, require high-quality instruction data with human annotations for SFT, which is both expensive and time-consuming. Furthermore, these methods require real-time LLM decoding during inference, which adds latency, particularly when generating relevance outputs for multiple items per query.

\section{Methodology} \label{sec3}

We introduce a new paradigm, E-CARE, that requires neither SFT, human annotation, nor real-time LLM decoding to bring commonsense reasoning knowledge into the e-commerce recommendation scenarios. To achieve this goal, we first design a pipeline to generate and refine a reasoning factor graph that captures commonsense reasoning knowledge underlying historical query-product interactions, and then train lightweight adapters on the graph's edges to efficiently predict reasoning factors for unseen queries and products as augmentations, thereby facilitating downstream recommendations. 

\begin{figure*}[t]
    \centering
\includegraphics[width=1\textwidth]{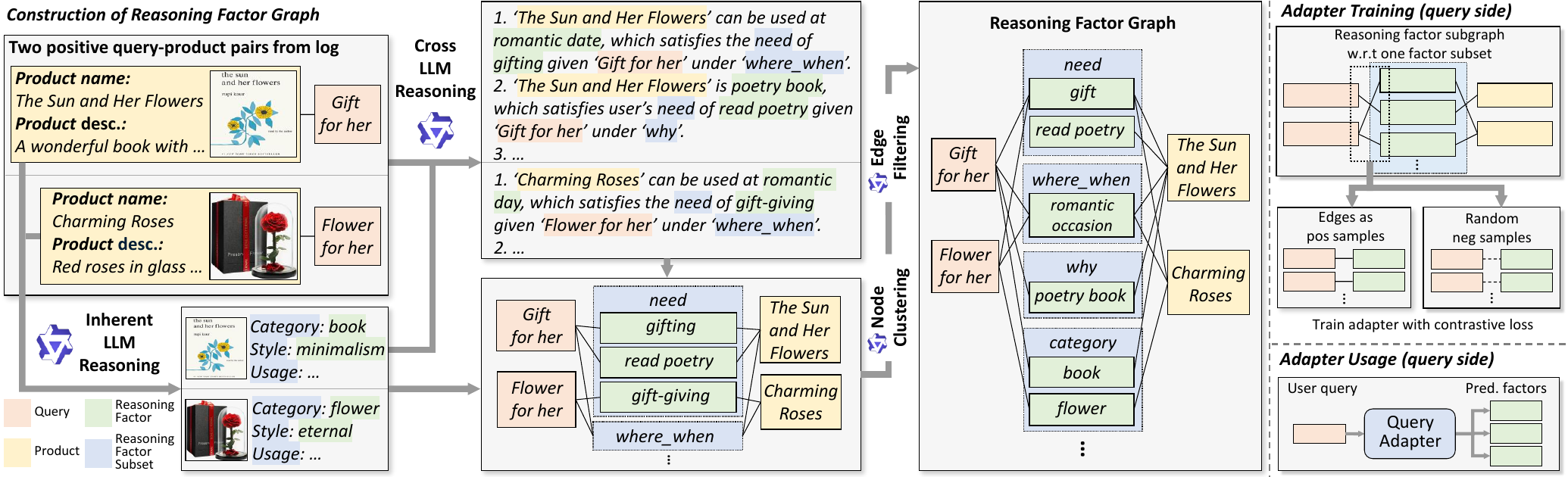}
    \caption{The overall diagram of E-CARE: The construction of the reasoning factor graph and adapter training/usage are shown on the left and right-hand side, respectively. Firstly, LLMs are prompted to perform commonsense reasoning over query-product historical pairs and product descriptions separately. The outputs are then parsed as the reasoning factor graph. Secondly, a node-clustering procedure is applied to merge nodes with similar semantic meanings across different reasoning factor subsets. Finally, LLM-based edge filtering is employed to remove unreliable edges from the reasoning factor graph, further improving quality. Given the graph, the adapter is trained with a contrastive loss, using graph edges as positive samples and random negatives. A trained adapter will be used to predict reasoning factors in the downstream application.}
    \label{fig:diagram_of_pipeline}
\end{figure*}

This section first outlines the pipeline for generating the reasoning factor graph, highlighting the specific designs we introduce to avoid the costs of SFT and human annotation. Then, introduce the adapter design and its training procedure. Figure \ref{fig:diagram_of_pipeline} provides an overview diagram.

\subsection{Construction of Reasoning Factor Graph} \label{sec:Construction_of_Reasoning_Factor_Graph}

Given a set of historical interactions $\sD$ containing samples of $(q, p)$ indicating that a user interacted with the product $p$ after inputting query $q$, where each product $p$ is associated with a text description $t_p$, we aim to generate a reasoning factor graph $\mathcal{G} = (\sQ, \sP, \sA, \sE)$ to represent the implicit factors between a pair of relevant query and product. Specifically, $\sQ:=\{q\mid (q,p)\in \sD\}$ and $\sP:=\{p\mid (q,p)\in\sD\}$ denote the set of query nodes and product nodes, respectively. $\sA$ denotes the set of nodes containing text segments as reasoning factors. $\sE$ is the set of edges that connect $\sQ$ and $\sP$ to $\sA$, and the commonly connected factors between a query and a product represent specific reasons why they are relevant.  

A three-stage pipeline comprising LLM reasoning, node clustering, and edge filtering is designed to mine reasoning factors and their connections from historical interactions by analyzing $\sD$ with LLMs.

\subsubsection{LLM Reasoning} \label{sec:LLM_reasoning}
In this stage, we aim to leverage LLMs' commonsense reasoning capabilities to generate the reasoning factors. We instruct LLMs to perform reasoning in 2 ways: \textit{Inherent LLM reasoning} to infer reasoning factors solely based on product descriptions and \textit{Cross LLM reasoning} to infer reasoning factors given historical query-product pairs.

\textbf{Inherent LLM reasoning.} 
Product descriptions contain valuable information that can facilitate recommendations, but they are often verbose and implicitly structured. To extract reasoning factors of a product, we leverage the DSPy framework \citep{khattab2023DSPyCompilingDeclarative} to generate prompts and perform guided commonsense reasoning conditioned on predefined reasoning types.

Specifically, we define a set of inherent reasoning types $\sF=\{f_i\}_{i=1}^{N}$, where each $f_i$ represents an aspect (e.g., ``usage'', ``style'') of a product that can be inferred from the product description. The complete descriptions of $\sF$ and an extraction example are provided in Appendix~\ref{appx:Product_Features_Extraction}.

For each product $p \in \sP$ with original textual description $t_p$, we extract a tuple of reasoning factors 
\begin{align}
    (t_p^{f_1}, \ldots, t_p^{f_N}),
\end{align}
where $t_p^{f_i}$ denotes the extracted text segment corresponding to reasoning type $f_i$. These extracted factors provide a structured representation of the product's inherent properties.

\textbf{Cross LLM reasoning.} 
Beyond product-inherent factors, we model the alignment between a query and a product by extracting cross-level reasoning factors from historical interactions $\sD$. 

Given a query–product pair $(q,p)$, we prompt the LLM to generate reasoning in a \emph{need–utility} format, where the latent \emph{need} $n$ implied by query $q$ is satisfied by the \emph{utility} $u$ provided by product $p$. A parser is then applied to extract the textual segments $n$ and $u$. We denote the structured interaction tuple as $(q,n,u,p)$ corresponding to the original pair $(q,p)$.

To capture diverse interaction semantics, we define a set of cross reasoning types $\sW=\{w_i\}_{i=1}^{M}$, where each $w_i$ specifies a semantic scope that characterizes the relationship between $q$ and $p$. Concretely, these reasoning types correspond to interrogative dimensions such as \textit{why} (underlying motivation or purpose), \textit{where\_when} (usage context, environment, temporal condition, or occasion), and \textit{who} (role or identity). Each reasoning type constrains the LLM to generate \emph{need–utility} explanations under a particular semantic perspective.

For each $(q,p)\in\sD$ and reasoning type $w \in \sW$, we obtain a structured interaction tuple $(q, n^{w}, u^{w}, p)$, where $(n^{w}, u^{w})$ denote the extracted cross reasoning factors conditioned on type $w$. Appendix~\ref{appx:LLM Commonsense Reasoning Prompt} provides the prompt template for the cross reasoning types \textit{who}.

After all samples in historical interactions $\sD$ have been analyzed, we collect all $t_p^{f_i}$, $n^w$, and $u^w$ into $\sA$ as the factor nodes. For each $(q,p)\in\sD$, we connect $q$ and $p$ to corresponding $t_p^{f_i}$ for $\forall f_i\in\sF$, as well as $n^w$ and $u^w$ for $\forall w\in\sW$. 

We further divide the factor nodes in $\sA$ into a collection of subsets $\sT$, where $\sA=\bigcup_{\sS\in\sT}\sS$, and each subset $\sS \subseteq \sA$. The division is primarily based on how factors are generated. Specifically, for each product inherent reasoning type $f_i\in\sF$, we collect all the factors related to $f_i$ as a subset. We include a subset of factors that are related to the utility $u^w$ for each cross reasoning type $w$. As for the need $n^w$, we put all needs with all cross reasoning types $w\in\sW$ as \emph{one} subset. This results in a reasoning factor graph $\mathcal{G}_0$ with number of subsets $|\sT|$ equals to $|\sF| + |\sW| + 1$.

\subsubsection{Node Clustering}\label{sec:node_clustering}

There are numerous similar factors in $\mathcal{G}_0$, so we conduct node clustering to shrink the size of $\mathcal{G}_0$. Within each subset $\sS\in\sT$, we conduct a clustering and aggregation to merge semantically similar factors, thereby reducing semantic redundancy between factors and enforcing denser correlations between queries and products. Specifically, we adopt a pretrained LLM, gte-Qwen2-7b-Instruct \citep{li2023GeneralTextEmbeddings}, to embed all the factors in a subset $\sS$. Then we adopt the clustering algorithm from \citet{reimers2019SentenceBERTSentenceEmbeddings} to divide factors in $\sS$ into clusters. For each cluster, we utilize LLMs to summarize the factors within it as a single new factor. The prompts can be found in Appendix \ref{Prompt_of_Edge_Clustering}. After merging the factors, we get a condensed reasoning factor graph $\mathcal{G}$.

\subsubsection{Edge Filtering}
We adopt pruning procedures to eliminate potential noisy edges. Inspired by LLM uncertainty evaluation \citep{ren2023SelfEvaluationImprovesSelective}, we design prompts that instruct LLMs to decide whether an edge $e$ is reasonable. The prompt templates are available in Appendix \ref{Edge_Filtering_Prompt}. Denote $s(e)$ as the prompt for $e$. We compute the confidence score of the edge via the contrastive probability score:
\begin{align}
    c_e = p(``YES"|s(e)) - p(``NO"| s(e))\,,
\end{align}
where $p(``YES"|s(e))$ is the probability of generating a ``YES'' token given the prompt $s(e)$ and $p(``NO"|s(e))$ is the probability of generating a ``NO'' token given the prompt $s(e)$. The subtraction serves as a calibration over the raw "YES" probability, thereby increasing the robustness of the contrastive probability score. Then, we keep only edges above a predefined threshold and use different thresholds for different edge types. Moreover, we also set the upper bound for the maximum number of factors that a query/product can connect with respect to each subset $\sS\in\sT$. 

The filtering procedure further reduces the size of the reasoning factor graph while improving its quality. After filtering, we obtain the clean, concise reasoning factor graph $\mathcal{G}$, which captures commonsense reasoning knowledge derived from historical interactions $\sD$ by LLMs. 

\subsection{Design of Adapters} \label{sec:Design of Adapters}
Lightweight adapters are trained to efficiently leverage commonsense knowledge from the reasoning factor graph $\mathcal{G}$ for unseen queries and products downstream tasks, thereby bypassing the need for LLMs to perform real-time decoding to acquire such knowledge during inference, as in previous work.

\subsubsection{Model of Query Adapter} \label{Model of Adapter}
The LLM-enhanced encoders project queries and factors into the latent space and select the top factors most closely aligned with the query. Specifically, for reasoning factors with a specific subset $\sS\in\sT$, and a query $q$, the encoder is designed as 

\begin{align}
    \text{enc}_\sS(q)=\text{MLP}(\text{LLM}(q)),
\end{align}
where $\text{LLM}(\cdot)$ denotes a function to map the query into an embedding via frozen LLMs \citep{li2023GeneralTextEmbeddings} and $\text{MLP}(\cdot)$ is a trainable multi-layer perceptron. Each reasoning factor node $f$ within the subset $\sS$ is encoded as
\begin{align}
    \text{enc}_\sS(f)=\text{MLP}(\text{LLM}(f)).
\end{align}
Then, the similarity between a query $q$ and a factor $f$ is estimated by
\begin{align}
    \text{sim}(q, f)=\frac{\langle \text{enc}_\sS(q), \text{enc}_\sS(f)\rangle}{\|\text{enc}_\sS(q)\|_2\cdot\|\text{enc}_\sS(f)\|_2}.
\end{align}

We take the top-$k$ factors within $\sS$ as the predicted factors of $q$. Eventually, we aggregate the predicted factors from all subsets into the predicted factors for $q$.

\subsubsection{Training the Query Adapter} \label{Training the Adapter}
We use the connections between query and factor nodes in $\mathcal{G}$ as labels to train the adapters. Specifically, give a factor subset $\sS$, we adopt the factors connected with query node $q\in\sQ$ as the set of positive labels
\begin{align}
    P^+_{\sS}(q) =\{n\in \sS\mid (q, n)\in \sE\}.
\end{align}
And we randomly sample a batch of factors from remaining nodes within $\sS$ as the negative labels 
\begin{align}
    P^-_{\sS}(q) =\{n\in \sS\mid (q, n)\notin \sE\}.
\end{align}
Then we apply InfoNCE loss \citep{oord2019RepresentationLearningContrastive} over those labels to train the query encoder of the factor subset $\sS$. 
\begin{align}
    & \mathcal{L}_\sS = \notag \\ 
    &\frac{1}{|\sQ|}\sum_{q\in\sQ} \frac{1}{|P^+_{\sS}(q)|}\sum_{n\in P^+_{\sS}(q)}\left[-\text{log}\frac{\text{exp}(\text{sim}(q, n))}{\sum_{m\in P^-_{\sS}(q) \cup \{n\}}\text{exp}(\text{sim}(q, m))}\right],
\end{align}
where $|\cdot|$ denotes the cardinality of a set.

For ease of notation, we denote $a(q)$ as the predicted set of factors for a query $q$ that is merged from the results of query adapters for each reasoning factor subset.

\subsubsection{Extension to Product Adapter}
\label{sec:product_adapter}

To handle cold-start products with few or no historical interactions, we use the same adapter design (\S \ref{Model of Adapter}) and training approach (\S \ref{Training the Adapter}) for training product adapters to predict reasoning factors given a product $p$. 

We denote $a(p)$ as the predicted set of factors for product $p$, computed by merging the results of product adapters for each reasoning factor subset. 

\begin{figure}[t]
    \centering
\includegraphics[width=1\columnwidth]{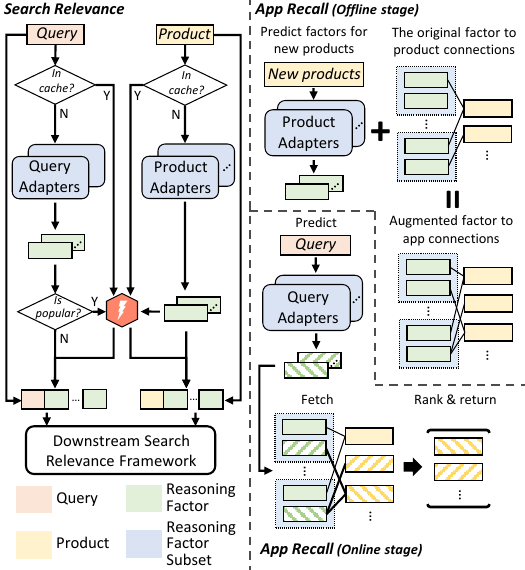}
    \caption{Overview diagram of two downstream applications integrated with the reasoning factor graph. \textit{Search Relevance}: The queries and products are augmented by trained adapters with reasoning factors before being fed into the downstream search relevance frameworks to predict search relevance scores. The reasoning factors for all products and popular queries are cached to accelerate inference speed. \textit{App Recall (Offline Inference)}: Product adapters are applied offline to connect new products to relevant factors. Such connections are then merged with the original factor to product connections on the graph to get the augmented factor to product connections \textit{App Recall (Online Inference)}: Query adapters are employed to predict root factors for the input query (The caching mechanism for popular queries is omitted in figure for simplicity), which are used to fetch products that are connected with root factors, In the end, the products are ranked and top k are returned.}
\label{fig:search_relevance_app_recall_pipeline}
\end{figure}

\section{Applications} \label{sec4}

In this section, we describe how to apply E-CARE to boost the performance of two downstream applications: search relevance (\S \ref{sec4.1}) and app recall (\S \ref{sec4.2}). The overview diagrams are shown in Figures \ref{fig:search_relevance_app_recall_pipeline}.

\subsection{Search Relevance} \label{sec4.1}

\subsubsection{Problem Statement}
A dataset $\sD^s$ consisting of samples of $(q, p, y)$ is given, where $q$ is a user query, $p$ is a product, and $y$ is a multi-class relevance label. Each $p$ is associated with a text description $t_p$. The goal is to predict relevance labels for unseen query-product pairs.

\subsubsection{Datasets}
We conduct experiments on 2 publicly available datasets, Amazon ESCI and WANDs, to assess the overall performance of our framework.

\begin{itemize}[leftmargin=*]
    \item \textbf{ESCI} \citep{reddy2022ShoppingQueriesDataset}: The dataset from KDD cup 2022, which provides manually labeled relevance judgments of query-product pairs from the e-commerce scenario. We conduct experiments on the English subset of Task 2, where each candidate item must be classified into one of four relevance labels: `Exact', `Substitute', `Complement', or `Irrelevant' (ESCI).
    \item \textbf{WANDs} \citep{Chen2022WANDSDatasetProduct}: The Wayfair ANnotation Dataset is a large-scale benchmark for e-commerce product search, where each query-product pair is categorized as `Exact', `Partial', or `Irrelevant'. We follow the train/dev/test split from a previous work \citep{cai2023ImprovingTextMatching}. The data splits can be found on the Huggingface hub \footnote{https://huggingface.co/datasets/napsternxg/wands}.
\end{itemize}
The detailed statistics of the dataset splits are shown in Table \ref{tab:dataset_stat}. 
\begin{table}[t]
    \centering
    \caption{Statistics of datasets.}
    \begin{tabular}{c|c|c|c} \hline
                            & ESCI (EN) & WANDs & Private dataset \\ \hline
    \# training samples     & 1,393,063 & 140,068 & 562,960 \\
    \# evaluation samples   & - & 46,690 & - \\
    \# test samples         & 425,762 & 46,690 & 200  \\
    \# unique queries       & 97,345 & 480 & 145,497 \\ 
    \# unique products      & 1,215,854 & 42,994 & 66,546 \\ \hline
    \end{tabular}
    \label{tab:dataset_stat}
\end{table}

\subsubsection{Baseline Frameworks}
We evaluate the performance of E-CARE over the following frameworks:
\begin{itemize}[leftmargin=*]
    \item \textbf{Bi-Encoder (BE)} \citep{reimers2019SentenceBERTSentenceEmbeddings}: This framework encodes query and product separately as embeddings and then feeds the combined embeddings into a prediction head (e.g., an MLP) to make a prediction on the relevant label. We use BERT-large-uncased \citep{devlinBERTPretrainingDeep}, DeBERTa-v3-large \citep{he2021debertav3}, and a frozen LLM, gte-Qwen2-7B-Instruct \citep{li2023GeneralTextEmbeddings}, as backbone encoders in our experiments.
    \item \textbf{Cross-Encoder (CE)} \citep{wu2022PracticeImprovingSearch}: In this framework, the query text and product description are concatenated and then encoded together, followed by a prediction head to predict the label. We apply BERT-large-uncased \citep{devlinBERTPretrainingDeep} and DeBERTa-v3-large \citep{he2021debertav3} as backbone encoders for this framework.
    \item \textbf{LLM Inference}: Following the previous work \citep{sun2023ChatGPTGoodSearch}, we prompt LLMs with few-shot examples \citep{brownLanguageModelsare} from the training set to directly make a prediction given a pair of queries and product as context. Llama-3.1-8b-Instruct \citep{grattafiori2024llama3herdmodels} is used as the backbone LLM in our experiment.
    \item \textbf{Ensemble}: An ensemble model combines prediction logits from multiple base encoders to achieve better performance than any single base encoder alone. We adopt the ensemble framework \citep{wu2022PracticeImprovingSearch}, ranked second on the KDD Cup 2022 leaderboard, in our experiment. The predictions of 3 backbone encoders (DeBERTa-v3-base \citep{he2021debertav3}, Big-Bird-base \citep{zaheer2021bigbirdtransformerslonger}, and CoCo-LM-base \citep{meng2021cocolmcorrectingcontrastingtext}) are aggregated by a LightGBM \citep{ke2017lightgbm} to make the final prediction.
\end{itemize}

\subsubsection{E-CARE Training} \label{sec.4.1.NURG Generation} 
First, the reasoning factor graph is constructed following the procedure described in \S \ref{sec:Construction_of_Reasoning_Factor_Graph} using only the positive subset $\sD^{s+}$ of the overall dataset $\sD^s$, where the labels of samples in the positive subset belong to the `Exact' class. We specify the cross reasoning types $\sW$ to be $\{\text{`where\_when'}, \text{`why', \text{`who'}}\}$ and the inherent reasoning types $\sF$ to be $\{\text{`category'}, \text{`style', \text{`usage'}}\}$ during graph generation and adapter training. The statistical analysis of the reasoning graph along the construction pipeline is reported in Appendix \ref{appx:Statistical Analysis}.

Second, query and product adapters are trained according to the procedure described in \S \ref{sec:Design of Adapters}. During the training of each adapter, we randomly split the corresponding query/product set into training and evaluation sets at a 9:1 ratio. We train the adapter on the training set and early stop based on performance on the evaluation set to avoid overfitting. The evaluation of the adapter training results is reported in Appendix \ref{Adapter Results Evaluation}.

\subsubsection{Augmentation with E-CARE}
We augment the product and query with the reasoning factor graph $\mathcal{G}$ and adapters. Specifically, for queries and products in the positive subset $\sD^{s+}$, we append the factors to which each query or product connects in $\mathcal{G}$ to the original queries and products. For other queries and products not in $\sD^{s+}$, we concatenate the predicted factors $a(q)$ and $a(p)$ with the original queries and products. Then we feed the augmented query and product into downstream search relevance frameworks for training and evaluation.

\subsubsection{Experiment Results}
Table \ref{tab:ESCI_WANDs} presents the results on the ESCI and WANDs datasets augmented by the E-CARE pipeline compared with the results of corresponding baseline models that do not include augmentation.

The experimental results on the ESCI dataset demonstrate the consistent benefits of incorporating reasoning factors into both bi-encoder and cross-encoder frameworks. For the bi-encoder, integrating commonsense factors into gte-Qwen2-7B-Instruct improves Macro F1 from 42.95 to 44.66 and Micro F1 from 67.81 to 68.37, indicating a stronger ability to distinguish related query-product pairs. A similar pattern is observed for DeBERTa-v3-large; the augmented variant achieves 48.70 Macro F1 and 68.27 Micro F1, again confirming that factor-based augmentation benefits the bi-encoder framework, which otherwise relies primarily on independent text representations. In contrast, cross-encoders achieve higher overall performance (e.g., 59.01 Macro F1 and 75.37 Micro F1 for DeBERTa-v3-large), but still benefit from additional factors, reaching 61.03 and 75.92, respectively. In addition, we evaluate the LLM inference setup using Llama-3.1-8B-Instruct, where factor-augmented inputs yield noticeable improvements, from 35.25 to 36.26 on Macro F1 and from 59.83 to 61.68 on Micro F1. This finding highlights that commonsense factors can benefit even generative models in discriminative inference settings. Finally, an ensemble combining DeBERTa-v3-base, BigBird-base, and CoCoLM-base further confirms the trend: the factor-augmented ensemble achieves 58.27 Macro F1 and 75.65 Micro F1, outperforming the non-augmented version (56.96/75.26).

On the WANDS dataset, a similar trend is observed, though the overall scores are substantially higher, leaving limited room for improvement. For bi-encoders, gte-Qwen2-7B-Instruct improves from 81.06 to 81.91 on Macro F1 and from 87.13 to 87.38 on Micro F1 after adding factors, and DeBERTa-v3-large exhibits a consistent improvement from 87.79 to 88.79 on Macro F1 and from 91.10 to 91.84 on Micro F1. However, BERT-large remains competitive, achieving 89.83 and 92.43 without augmentation. Cross-encoders outperform bi-encoders, with DeBERTa-v3-large achieving 91.39 Macro F1 and 93.38 Micro F1, demonstrating that while base architectures already capture strong cross-interactions, commonsense reasoning signals can still yield marginal yet stable gains.

Taken together, the experiment results across the ESCI and WANDS datasets validate the effectiveness of the E-CARE augmentations. Moreover, a comparison between the two datasets indicates that the commonsense reasoning factors generated by our pipeline are most impactful in more challenging and diverse environments (e.g., Amazon ESCI), where ambiguity and semantic variability are prevalent. In contrast, on relatively homogeneous datasets such as WANDS, the effect is less pronounced because of the limited diversity of query-product relations.

\begin{table}[t]
    \centering
    \caption{
    Search relevance results on the ESCI and WANDs dataset. `w/ fts' indicates that the query and product are augmented with predicted factors before being fed into the downstream frameworks.}
    \resizebox{\columnwidth}{!}{%
    \begin{tabular}{c|c|>{\centering\arraybackslash}p{0.7cm}|>{\centering\arraybackslash}p{0.9cm}>{\centering\arraybackslash}p{0.9cm}>{\centering\arraybackslash}p{0.9cm}>{\centering\arraybackslash}p{0.9cm}}
    \hline
    \multirow{2}{*}{Framework} & \multirow{2}{*}{Backbone Model} & \multirow{2}{*}{w/ fts} & \multicolumn{2}{c}{ESCI} & \multicolumn{2}{c}{WANDs} \\ 
    \cline{4-7} & & & Mac F1 & Mic F1 & Mac F1 & Mic F1 \\ \hline
    \multirow{6}{*}{BE}             & \multirow{2}{*}{gte-Qwen2-7B}     & \xmark & 42.95 & 67.81 & 81.06 & 87.13 \\ \cline{3-3}
                                    &                                   & \cmark & \textbf{44.66} & \textbf{68.37} & \textbf{81.91} & \textbf{87.38}  \\ \cline{2-7}
                                    & \multirow{2}{*}{BERT-large}             & \xmark & 48.98 & \textbf{69.77} & \textbf{89.83} & \textbf{92.43} \\ \cline{3-3}
                                    &                                   & \cmark & \textbf{49.71} & 68.57 & 89.54 & 92.32 \\ \cline{2-7}
                                    & \multirow{2}{*}{DeBERTa-v3-large}       & \xmark & 46.87 & 67.25 & 87.79 & 91.10  \\ \cline{3-3}
                                    &                                   & \cmark & \textbf{48.70} & \textbf{68.27}& \textbf{88.79} & \textbf{91.84} \\ \hline
    \multirow{4}{*}{CE}             & \multirow{2}{*}{BERT-large}             & \xmark & 55.82 & 73.36 & \textbf{90.55} & \textbf{92.77} \\ \cline{3-3}
                                    &                                   & \cmark & \textbf{57.61} & \textbf{74.01} & 90.30 & 92.66 \\ \cline{2-7}
                                    & \multirow{2}{*}{DeBERTa-v3-large}       & \xmark & 59.01 & 75.37 & 91.33 & 93.28 \\ \cline{3-3}
                                    &                                   & \cmark & \textbf{61.03} & \textbf{75.92} & \textbf{91.39} & \textbf{93.38}  \\ \hline
    \multirow{2}{*}{\shortstack{LLM\\Inference}}     & \multirow{2}{*}{Llama-3.1-8B}     & \xmark & 35.25 & 59.83 & 37.06 & 39.17 \\ \cline{3-3}
                                    &                                   & \cmark & \textbf{36.26}  & \textbf{61.68} & \textbf{49.85} & \textbf{51.50} \\ \hline
    \multirow{4}{*}{Ensemble}       & \multirow{4}{*}{\shortstack{DeBERTa-v3-base\\BigBird-base\\CoCoLM-base}} & \multirow{2}{*}{\xmark} & \multirow{2}{*}{56.96} & \multirow{2}{*}{75.26} & \multirow{2}{*}{-} & \multirow{2}{*}{-} \\
                                    &                                   & & & & & \\ \cline{3-3}
                                    &                                   & \multirow{2}{*}{\cmark} & \multirow{2}{*}{\textbf{58.27}} & \multirow{2}{*}{\textbf{75.65}} & \multirow{2}{*}{-} & \multirow{2}{*}{-} \\ 
                                    &                                   & & & & & \\ \hline
    \end{tabular}
    }
    \label{tab:ESCI_WANDs}
\end{table}

\subsubsection{Case study}
We conduct case studies to demonstrate how our pipeline predicts relevance labels. The case studies are shown in Table \ref{tab:case_study}.

In the first example, the model fails to match the term `enhancement' to `energy supplement' using only the original query and product. But with predicted category (i.e., `supplements'), purpose (i.e., `energy support'), and user profiles (i.e., `health-conscious individuals') from our pipeline, the similarity of query and product becomes closer, so that the model can predict the right label.

In the second example, the query is vague and does not specify the exact product type, resulting in an incorrect prediction. With E-CARE augmentation injected, the backbone model can discover similarity between intentions (i.e., `natural care' and `digestive health'), features (i.e., `plant-based' and `vegan'), and user tendencies (i.e., `health-conscious individuals'), resulting in a correct prediction.

\begin{table*}[h]
    \centering
    \caption{Case studies on samples from the ESCI dataset. The text in double quotes is the original query and product title. In contrast, the text marked in different colors is the information E-CARE provides that helps turn the prediction from incorrect to correct. Note that detailed product descriptions have been omitted for brevity purposes.}
    \resizebox{\textwidth}{!}{
    \begin{tabular}{>{\arraybackslash}p{7cm}|>{\arraybackslash}p{7cm}|>{\centering\arraybackslash}p{1.6cm}|>{\centering\arraybackslash}p{1.6cm}|>{\centering\arraybackslash}p{1.6cm}}
    \hline
    \multirow{2}{*}{Query} & \multirow{2}{*}{Product} & \multirow{2}{*}{\shortstack{Prediction\\w/o E-CARE}} & \multirow{2}{*}{\shortstack{Prediction\\w/ E-CARE}} & \multirow{2}{*}{\shortstack{Ground\\Truth}} \\
    &  &  &  &   \\ \hline
    \textit{``100\% all natural male enhancement without caffeine''} \newline\newline belongs to categories of [\textcolor{violet}{supplements}, personal care], has style of [herbal, plant-based], can be used for [supplement, \textcolor{red}{energy support}], with intention of [health support, supplementation], can be used at [at home, daily], can be used by [adults, \textcolor{blue}{health-conscious individuals}], with purpose of [supplement, dietary supplement] & \textit{``Rise Up, Red Edition Natural Energy \textcolor{violet}{supplement}''} \newline\newline belongs to categories of [supplements], has style of [Red Edition], can be used for [energy boost, pre-workout], with intention of [\textcolor{red}{energy booster}, supplementation], can be used at [workout, gym], can be used by [athletes, \textcolor{blue}{health-conscious individuals}], with purpose of [energy, supplement]  & Irrelevant & Exact & Exact  \\ \hline
    \textit{``100\% organic, pure without any mix.''} \newline\newline belongs to categories of [food, beverages], has style of [natural, \textcolor{violet}{plant-based}], can be used for [aromatherapy, facial care], with intention of [\textcolor{blue}{natural remedy, natural care}], can be used at [kitchen, home], can be used by [\textcolor{red}{health-conscious individuals}, wellness enthusiast], with purpose of [natural, plant-based] & \textit{``Garden of Life Raw Organic Protein Vanilla Powder: Certified Vegan, Gluten Free, Organic''} \newline\newline belongs to categories of [protein powder], has style of [vanilla], can be used for [post-workout, \textcolor{blue}{digestive health}], with intention of [protein amino intake, protein], can be used at [home, gym], can be used by [vegans, \textcolor{red}{health-conscious individuals}], with purpose of [whey protein powder, \textcolor{violet}{vegan}] & Irrelevant & Exact & Exact   \\ \hline
    
    \end{tabular}}
    \label{tab:case_study}
\end{table*}

\subsection{App Recall} \label{sec4.2}

\subsubsection{Problem Statement}

A dataset $\sD^r$ consisting of samples of $(q, p)$ is provided, where $q$ is the user query and $p$ is an app with a text-based description $t_p$ that the user has interacted with. All apps are from a pre-defined app set $\sP$. The goal is to recall a subset of apps from $\sP$ that the user may be interested in, given any input query $q$.


\subsubsection{Datasets}
We conduct app recall experiments on a private dataset from our app recommendation scenario. The anonymous user search logs are used as the training dataset, with each sample containing a query-app pair, representing that the user clicks the app after submitting the query. A human-annotated dataset is provided as the test set, comprising 200 queries and, for each query, a corresponding list of apps that satisfy it.

The detailed statistics of the private dataset are shown in Table~\ref{tab:dataset_stat}.

\subsubsection{Baselines}
We compare our method with our current app recall system, which incorporates, but is not limited to, recall results from multiple strategies, including keyword similarity, semantic similarity, and popularity-based recommendation.

\subsubsection{E-CARE Training}
The construction of the reasoning factor graph and the adapter training procedure for the app recall scenario are similar to those described in \S \ref{sec.4.1.NURG Generation}. The only difference is that the construction of the reasoning factor graph is based on the overall dataset $\sD^r$, since all samples are treated as positive samples. The E-CARE training is scheduled to be conducted on a monthly basis.

\subsubsection{App Recall with E-CARE}
First, in the offline stage, the trained product adapters are employed to predict reasoning factors for unseen apps outside the training set but within the app candidate set for the day. The predicted connections are then merged with existing factor-to-product edges $\sE_{f,q}$ on the reasoning factor graph $\mathcal{G}$ to acquire augmented factor-to-product edges $\sE^\prime_{f,q}$. The offline stage is scheduled to be conducted on a daily basis.

Second, in the online stage, given a user query $q$, we obtain the set of connected factors via the query adapters, denoted $a(q)$, and further acquire a set of apps, $\sP^{\prime}$, such that 

\begin{align}
    \sP^{\prime} 
    = \left\{ p \in \sP \;\middle|\;\exists f \in a(q),\; (f,p) \in \sE^{\prime}_{f,q}\right\}.
\end{align}

Then, we use the count of overlapped factors between $q$ and any app $p \in \sP^\prime$ as a similarity measure $\text{sim}(q,p)$, i.e.,
\begin{align}
    \text{sim}(q,p) = \left|\{(f, p)\mid f\in a(q), p\in \sP^\prime, (f, p)\in \sE^\prime_{f,q}\}\right|,
\end{align}
where $|\cdot|$ denotes the cardinality of a set. We then select the top-$k$ apps with the highest similarity scores as the final recall app list to query $q$.

\subsubsection{Experiment Results}
Table \ref{tab:private} presents the offline evaluation results for the E-CARE pipeline compared with the baseline system on the private dataset. Our method substantially outperforms the existing online recall system across both evaluation metrics. Specifically, Recall@5 improves from 51.3\% to 62.4\%, while Precision@5 increases from 41.0\% to 53.1\%. These gains of over 10 percentage points in both recall and precision highlight the effectiveness of the proposed framework in retrieving more relevant candidate apps while reducing noise in the top-ranked results. 

Moreover, the E-CARE pipeline has already been deployed in the recall stage of our browser app advertisement system to retrieve relevant app candidates from a user query in the search bar. In the A/B test with 20\% traffic, total Value Per Mille (VPM), Click-Through Rate (CTR), and Conversion Rate (CVR) increased by 1.41\%, 0.65\%, and 3.39\%, respectively. The online results suggest that our method better captures the underlying connections between user queries and apps than the baselines by leveraging a commonsense reasoning signal, which ultimately yields higher revenue for both advertisers and our platform.

\begin{table}[t]
    \centering
    \caption{
    Item recall results on the private dataset.
    }
    \begin{tabular}{c|c|c}
    \hline
    \multirow{2}{*}{Method} & \multicolumn{2}{c}{private dataset} \\ 
    \cline{2-3} & Recall@5 & Precision@5 \\ \hline
    online recall system      & 51.3 & 41.0 \\  
    ECARE-based recall         & 62.4 & 53.1 \\ \hline
    \end{tabular}
    \label{tab:private}
\end{table}

\section{Conclusion} \label{conclusion}

In this work, we propose E-CARE, an Efficient Commonsense Augmented Recommendation Enhancer, which constructs a reasoning factor graph from historical query-product interactions without requiring supervised fine-tuning or human annotations. By leveraging LLMs for commonsense reasoning generation, node clustering, and edge filtering, E-CARE produces a high-quality reasoning factor graph that captures meaningful connections between queries and products. Our experiments demonstrate that the resulting reasoning factor graph effectively enhances downstream tasks, such as search relevance and app recall, while avoiding the cost and latency associated with real-time LLM decoding. These findings highlight the potential of E-CARE as a scalable and efficient framework for incorporating commonsense reasoning signals into recommendation systems.

\bibliographystyle{ACM-Reference-Format}

\bibliography{nurg_truncate}  

@inproceedings{brownLanguageModelsare,
author = {Brown, Tom B. and Mann, Benjamin and et al},
title = {Language models are few-shot learners},
year = {2020},
isbn = {9781713829546},
publisher = {Curran Associates Inc.},
address = {Red Hook, NY, USA},
booktitle = {Proceedings of the 34th International Conference on Neural Information Processing Systems},
articleno = {159},
numpages = {25},
location = {Vancouver, BC, Canada},
series = {NIPS '20}
}

@misc{cai2023ImprovingTextMatching,
  title = {Improving {{Text Matching}} in {{E-Commerce Search}} with {{A Rationalizable}}, {{Intervenable}} and {{Fast Entity-Based Relevance Model}}},
  author = {Cai, Jiong and Jiang, Yong and et al},
  year = {2023},
  number = {arXiv:2307.00370},
  eprint = {2307.00370},
  primaryclass = {cs},
  publisher = {arXiv},
  langid = {english}
}

@inproceedings{dai2019DeeperTextUnderstanding,
  title = {Deeper {{Text Understanding}} for {{IR}} with {{Contextual Neural Language Modeling}}},
  booktitle = {Proceedings of the 42nd {{International ACM SIGIR Conference}} on {{Research}} and {{Development}} in {{Information Retrieval}}},
  author = {Dai, Zhuyun and Callan, Jamie},
  year = {2019},
  publisher = {ACM},
  copyright = {https://www.acm.org/publications/policies/copyright\_policy\#Background},
  langid = {english}
}

@inproceedings{devlinBERTPretrainingDeep,
    title = "{BERT}: Pre-training of Deep Bidirectional Transformers for Language Understanding",
    author = "Devlin, Jacob  and
      Chang, Ming-Wei  and
      et al",
    editor = "Burstein, Jill  and
      Doran, Christy  and
      Solorio, Thamar",
    booktitle = "Proceedings of the 2019 Conference of the North {A}merican Chapter of the Association for Computational Linguistics: Human Language Technologies, Volume 1 (Long and Short Papers)",
    month = jun,
    year = "2019",
    address = "Minneapolis, Minnesota",
    publisher = "Association for Computational Linguistics",
    url = "https://aclanthology.org/N19-1423/",
    doi = "10.18653/v1/N19-1423",
    pages = "4171--4186",
}

@inproceedings{hofstatter2022IntroducingNeuralBag,
  title = {Introducing {{Neural Bag}} of {{Whole-Words}} with {{ColBERTer}}: {{Contextualized Late Interactions}} Using {{Enhanced Reduction}}},
  shorttitle = {Introducing {{Neural Bag}} of {{Whole-Words}} with {{ColBERTer}}},
  booktitle = {Proceedings of the 31st {{ACM International Conference}} on {{Information}} \& {{Knowledge Management}}},
  author = {Hofst{\"a}tter, Sebastian and Khattab, Omar and et al},
  year = {2022},
  publisher = {ACM},
  copyright = {https://creativecommons.org/licenses/by/4.0/},
  langid = {english}
}

@inproceedings{huang2013Learningdeepstructured,
  title = {Learning Deep Structured Semantic Models for Web Search Using Clickthrough Data},
  booktitle = {Proceedings of the 22nd {{ACM}} International Conference on {{Information}} \& {{Knowledge Management}}},
  author = {Huang, Po-Sen and He, Xiaodong and et al},
  year = {2013},
  publisher = {ACM},
  copyright = {https://www.acm.org/publications/policies/copyright\_policy\#Background},
  langid = {english}
}

@inproceedings{
humeau2020Polyencodersarchitecturespretraining,
title={Poly-encoders: Architectures and Pre-training Strategies for Fast and Accurate Multi-sentence Scoring},
author={Samuel Humeau and Kurt Shuster and et al},
booktitle={International Conference on Learning Representations},
year={2020},
url={https://openreview.net/forum?id=SkxgnnNFvH}
}

@inproceedings{karpukhin2020DensePassageRetrievala,
  title = {Dense {{Passage Retrieval}} for {{Open-Domain Question Answering}}},
  booktitle = {Proceedings of the 2020 {{Conference}} on {{Empirical Methods}} in {{Natural Language Processing}} ({{EMNLP}})},
  author = {Karpukhin, Vladimir and Oguz, Barlas and et al},
  year = {2020},
  publisher = {Association for Computational Linguistics},
  langid = {english}
}

@inproceedings{keenOkapiTREC3,
  title={Okapi at TREC-3},
  author={Stephen E. Robertson and Steve Walker and et al},
  booktitle={Text Retrieval Conference},
  year={1994},
  url={https://api.semanticscholar.org/CorpusID:41563977}
}

@inproceedings{khattab2020ColBERTEfficientEffective,
  title = {{{ColBERT}}: {{Efficient}} and {{Effective Passage Search}} via {{Contextualized Late Interaction}} over {{BERT}}},
  shorttitle = {{{ColBERT}}},
  booktitle = {Proceedings of the 43rd {{International ACM SIGIR Conference}} on {{Research}} and {{Development}} in {{Information Retrieval}}},
  author = {Khattab, Omar and Zaharia, Matei},
  year = {2020},
  publisher = {ACM},
  copyright = {https://www.acm.org/publications/policies/copyright\_policy\#Background},
  langid = {english}
}

@inproceedings{
khattab2023DSPyCompilingDeclarative,
title={{DSP}y: Compiling Declarative Language Model Calls into State-of-the-Art Pipelines},
author={Omar Khattab and Arnav Singhvi and et al},
booktitle={The Twelfth International Conference on Learning Representations},
year={2024},
url={https://openreview.net/forum?id=sY5N0zY5Od}
}

@inproceedings{kojima2023LargeLanguageModels,
author = {Kojima, Takeshi and Gu, Shixiang Shane and et al},
title = {Large language models are zero-shot reasoners},
year = {2022},
isbn = {9781713871088},
publisher = {Curran Associates Inc.},
address = {Red Hook, NY, USA},
booktitle = {Proceedings of the 36th International Conference on Neural Information Processing Systems},
articleno = {1613},
numpages = {15},
location = {New Orleans, LA, USA},
series = {NIPS '22}
}

@inproceedings{kong2022MultiAspectDenseRetrievala,
  title = {Multi-{{Aspect Dense Retrieval}}},
  booktitle = {Proceedings of the 28th {{ACM SIGKDD Conference}} on {{Knowledge Discovery}} and {{Data Mining}}},
  author = {Kong, Weize and Khadanga, Swaraj and et al},
  year = {2022},
  publisher = {ACM},
  copyright = {https://creativecommons.org/licenses/by/4.0/},
  langid = {english}
}

@misc{kumar2021NeuralSearchLearning,
  title = {Neural {{Search}}: {{Learning Query}} and {{Product Representations}} in {{Fashion E-commerce}}},
  shorttitle = {Neural {{Search}}},
  author = {Kumar, Lakshya and Sarkar, Sagnik},
  year = {2021},
  number = {arXiv:2107.08291},
  eprint = {2107.08291},
  primaryclass = {cs},
  publisher = {arXiv},
  langid = {english}
}

@inproceedings{li2021MoreRobustDense,
  title = {More {{Robust Dense Retrieval}} with {{Contrastive Dual Learning}}},
  booktitle = {Proceedings of the 2021 {{ACM SIGIR International Conference}} on {{Theory}} of {{Information Retrieval}}},
  author = {Li, Yizhi and Liu, Zhenghao and and et al},
  year = {2021},
  publisher = {ACM},
  copyright = {https://www.acm.org/publications/policies/copyright\_policy\#Background},
  langid = {english}
}

@inproceedings{li2023EcomGPTInstructiontuningLarge,
author = {Li, Yangning and Ma, Shirong and et al},
title = {EcomGPT: instruction-tuning large language models with chain-of-task tasks for E-commerce},
year = {2024},
isbn = {978-1-57735-887-9},
publisher = {AAAI Press},
url = {https://doi.org/10.1609/aaai.v38i17.29820},
doi = {10.1609/aaai.v38i17.29820},
booktitle = {Proceedings of the Thirty-Eighth AAAI Conference on Artificial Intelligence and Thirty-Sixth Conference on Innovative Applications of Artificial Intelligence and Fourteenth Symposium on Educational Advances in Artificial Intelligence},
articleno = {2072},
numpages = {9},
series = {AAAI'24/IAAI'24/EAAI'24}
}

@inproceedings{li2023StructureAwareLanguageModela,
  title = {Structure-{{Aware Language Model Pretraining Improves Dense Retrieval}} on {{Structured Data}}},
  booktitle = {Findings of the {{Association}} for {{Computational Linguistics}}: {{ACL}} 2023},
  author = {Li, Xinze and Liu, Zhenghao and et al},
  year = {2023},
  publisher = {Association for Computational Linguistics},
  langid = {english}
}

@inproceedings{lin2021InBatchNegativesKnowledge,
  title = {In-{{Batch Negatives}} for {{Knowledge Distillation}} with {{Tightly-Coupled Teachers}} for {{Dense Retrieval}}},
  booktitle = {Proceedings of the 6th {{Workshop}} on {{Representation Learning}} for {{NLP}} ({{RepL4NLP-2021}})},
  author = {Lin, Sheng-Chieh and Yang, Jheng-Hong and et al},
  year = {2021},
  publisher = {Association for Computational Linguistics},
  langid = {english}
}

@article{liu2025RaCTRankingawareChainofThought,
  title={RaCT: Ranking-aware Chain-of-Thought Optimization for LLMs},
  author={Haowei Liu and Xuyang Wu and Guohao Sun and Hsin-Tai Wu and Zhiqiang Tao and Yi Fang},
  journal={Proceedings of the 2025 Annual International ACM SIGIR Conference on Research and Development in Information Retrieval in the Asia Pacific Region},
  year={2024},
  url={https://api.semanticscholar.org/CorpusID:274860079}
}

@misc{ma2023ZeroShotListwiseDocument,
  title = {Zero-{{Shot Listwise Document Reranking}} with a {{Large Language Model}}},
  author = {Ma, Xueguang and Zhang, Xinyu and et al},
  year = {2023},
  number = {arXiv:2305.02156},
  eprint = {2305.02156},
  primaryclass = {cs},
  publisher = {arXiv},
  langid = {english}
}

@inproceedings{nogueira2020DocumentRankingPretrained,
  title = {Document {{Ranking}} with a {{Pretrained Sequence-to-Sequence Model}}},
  booktitle = {Findings of the {{Association}} for {{Computational Linguistics}}: {{EMNLP}} 2020},
  author = {Nogueira, Rodrigo and Jiang, Zhiying and et al},
  year = {2020},
  publisher = {Association for Computational Linguistics},
  langid = {english}
}

@misc{nogueira2020PassageRerankingBERT,
  title = {Passage {{Re-ranking}} with {{BERT}}},
  author = {Nogueira, Rodrigo and Cho, Kyunghyun},
  year = {2020},
  number = {arXiv:1901.04085},
  eprint = {1901.04085},
  primaryclass = {cs},
  publisher = {arXiv},
  langid = {english}
}

@misc{oord2019RepresentationLearningContrastive,
  title = {Representation {{Learning}} with {{Contrastive Predictive Coding}}},
  author = {van den Oord, Aaron and Li, Yazhe and et al},
  year = {2019},
  number = {arXiv:1807.03748},
  eprint = {1807.03748},
  primaryclass = {cs},
  publisher = {arXiv},
  langid = {english}
}

@misc{openai2024GPT4TechnicalReporta,
  title = {{{GPT-4 Technical Report}}},
  author = {OpenAI and Achiam, Josh and Adler, Steven and et al},
  year = {2024},
  number = {arXiv:2303.08774},
  eprint = {2303.08774},
  primaryclass = {cs},
  publisher = {arXiv},
  langid = {english}
}

@misc{openai2024OpenAIo1System,
  title = {{{OpenAI}} O1 {{System Card}}},
  author = {OpenAI and Jaech, Aaron and Kalai, Adam and et al},
  year = {2024},
  number = {arXiv:2412.16720},
  eprint = {2412.16720},
  primaryclass = {cs},
  publisher = {arXiv},
  langid = {english}
}

@misc{pradeep2023RankVicunaZeroShotListwisea,
  title = {{{RankVicuna}}: {{Zero-Shot Listwise Document Reranking}} with {{Open-Source Large Language Models}}},
  shorttitle = {{{RankVicuna}}},
  author = {Pradeep, Ronak and Sharifymoghaddam, Sahel and et al},
  year = {2023},
  number = {arXiv:2309.15088},
  eprint = {2309.15088},
  primaryclass = {cs},
  publisher = {arXiv},
  langid = {english}
}

@inproceedings{qin2024LargeLanguageModels,
  title = {Large {{Language Models}} Are {{Effective Text Rankers}} with {{Pairwise Ranking Prompting}}},
  booktitle = {Findings of the {{Association}} for {{Computational Linguistics}}: {{NAACL}} 2024},
  author = {Qin, Zhen and Jagerman, Rolf and et al},
  year = {2024},
  publisher = {Association for Computational Linguistics},
  langid = {english}
}

@inproceedings{qin2025TongSearchQRReinforcedQuery,
    title = "Reinforced Query Reasoners for Reasoning-intensive Retrieval Tasks",
    author = "Qin, Xubo  and
      Bai, Jun  and
      Li, Jiaqi  and
      Jia, Zixia  and
      Zheng, Zilong",
    editor = "Christodoulopoulos, Christos  and
      Chakraborty, Tanmoy  and
      Rose, Carolyn  and
      Peng, Violet",
    booktitle = "Proceedings of the 2025 Conference on Empirical Methods in Natural Language Processing",
    month = nov,
    year = "2025",
    address = "Suzhou, China",
    publisher = "Association for Computational Linguistics",
    url = "https://aclanthology.org/2025.emnlp-main.1078/",
    doi = "10.18653/v1/2025.emnlp-main.1078",
    pages = "21250--21263",
    ISBN = "979-8-89176-332-6",
}

@inproceedings{
rafailovDirectPreferenceOptimization,
title={Direct Preference Optimization: Your Language Model is Secretly a Reward Model},
author={Rafael Rafailov and Archit Sharma and et al},
booktitle={Thirty-seventh Conference on Neural Information Processing Systems},
year={2023},
url={https://openreview.net/forum?id=HPuSIXJaa9}
}

@article{raffelExploringLimitsTransfer,
author = {Raffel, Colin and Shazeer, Noam and et al},
title = {Exploring the limits of transfer learning with a unified text-to-text transformer},
year = {2020},
issue_date = {January 2020},
publisher = {JMLR.org},
volume = {21},
number = {1},
issn = {1532-4435},
journal = {J. Mach. Learn. Res.},
month = jan,
articleno = {140},
numpages = {67}
}

@inproceedings{ramosUsingTFIDFDetermine,
  title={Using TF-IDF to Determine Word Relevance in Document Queries},
  author={Juan Enrique Ramos},
  year={2003},
  url={https://api.semanticscholar.org/CorpusID:14638345}
}

@misc{reddy2022ShoppingQueriesDataset,
  title = {Shopping {{Queries Dataset}}: {{A Large-Scale ESCI Benchmark}} for {{Improving Product Search}}},
  shorttitle = {Shopping {{Queries Dataset}}},
  author = {Reddy, Chandan K. and M{\`a}rquez, Llu{\'i}s and et al},
  year = {2022},
  number = {arXiv:2206.06588},
  eprint = {2206.06588},
  primaryclass = {cs},
  publisher = {arXiv},
  langid = {english}
}

@inproceedings{reimers2019SentenceBERTSentenceEmbeddings,
    title = "Sentence-{BERT}: Sentence Embeddings using {S}iamese {BERT}-Networks",
    author = "Reimers, Nils  and
      Gurevych, Iryna",
    editor = "Inui, Kentaro  and
      Jiang, Jing  and
      Ng, Vincent  and
      Wan, Xiaojun",
    booktitle = "Proceedings of the 2019 Conference on Empirical Methods in Natural Language Processing and the 9th International Joint Conference on Natural Language Processing (EMNLP-IJCNLP)",
    month = nov,
    year = "2019",
    address = "Hong Kong, China",
    publisher = "Association for Computational Linguistics",
    url = "https://aclanthology.org/D19-1410/",
    doi = "10.18653/v1/D19-1410",
    pages = "3982--3992",
}

@InProceedings{ren2023SelfEvaluationImprovesSelective,
  title = {Self-Evaluation Improves Selective Generation in Large Language Models},
  author = {Ren, Jie and Zhao, Yao and et al},
  booktitle = {Proceedings on "I Can't Believe It's Not Better: Failure  Modes in the Age of Foundation Models" at NeurIPS 2023 Workshops},
  pages = {49--64},
  year = {2023},
  editor = {Antorán, Javier and Blaas, Arno and Buchanan, Kelly and Feng, Fan and Fortuin, Vincent and Ghalebikesabi, Sahra and Kriegler, Andreas and Mason, Ian and Rohde, David and Ruiz, Francisco J. R. and Uelwer, Tobias and Xie, Yubin and Yang, Rui},
  volume = {239},
  series = {Proceedings of Machine Learning Research},
  month = {16 Dec},
  publisher = {PMLR},
  url = {https://proceedings.mlr.press/v239/ren23a.html},
}

@inproceedings{sachan2022ImprovingPassageRetrieval,
  title = {Improving {{Passage Retrieval}} with {{Zero-Shot Question Generation}}},
  booktitle = {Proceedings of the 2022 {{Conference}} on {{Empirical Methods}} in {{Natural Language Processing}}},
  author = {Sachan, Devendra and Lewis, Mike and et al},
  year = {2022},
  publisher = {Association for Computational Linguistics},
  langid = {english}
}

@inproceedings{santhanam2022ColBERTv2EffectiveEfficient,
  title = {{{ColBERTv2}}: {{Effective}} and {{Efficient Retrieval}} via {{Lightweight Late Interaction}}},
  shorttitle = {{{ColBERTv2}}},
  booktitle = {Proceedings of the 2022 {{Conference}} of the {{North American Chapter}} of the {{Association}} for {{Computational Linguistics}}: {{Human Language Technologies}}},
  author = {Santhanam, Keshav and Khattab, Omar and et al},
  year = {2022},
  publisher = {Association for Computational Linguistics},
  langid = {english}
}

@inproceedings{schlatt2024InvestigatingEffectsSparse,
author = {Schlatt, Ferdinand and Fr\"{o}be, Maik and et al},
title = {Investigating the\&nbsp;Effects of\&nbsp;Sparse Attention on\&nbsp;Cross-Encoders},
year = {2024},
isbn = {978-3-031-56026-2},
publisher = {Springer-Verlag},
address = {Berlin, Heidelberg},
url = {https://doi.org/10.1007/978-3-031-56027-9_11},
doi = {10.1007/978-3-031-56027-9_11},
booktitle = {Advances in Information Retrieval: 46th European Conference on Information Retrieval, ECIR 2024, Glasgow, UK, March 24–28, 2024, Proceedings, Part I},
pages = {173–190},
numpages = {18},
location = {Glasgow, United Kingdom}
}

@inproceedings{shan2023TwoTowerAttributeGuidedb,
  title = {Beyond {{Two-Tower}}: {{Attribute Guided Representation Learning}} for {{Candidate Retrieval}}},
  shorttitle = {Beyond {{Two-Tower}}},
  booktitle = {Proceedings of the {{ACM Web Conference}} 2023},
  author = {Shan, Hongyu and Zhang, Qishen and et al},
  year = {2023},
  publisher = {ACM},
  copyright = {https://www.acm.org/publications/policies/copyright\_policy\#Background},
  langid = {english}
}

@inproceedings{shang2025KnowledgeDistillationEnhancing,
author = {Shang, Hongwei and Vo, Nguyen and et al},
title = {Knowledge Distillation for Enhancing Walmart E-commerce Search Relevance Using Large Language Models},
year = {2025},
isbn = {9798400713316},
publisher = {Association for Computing Machinery},
address = {New York, NY, USA},
url = {https://doi.org/10.1145/3701716.3715242},
doi = {10.1145/3701716.3715242},
booktitle = {Companion Proceedings of the ACM on Web Conference 2025},
pages = {449–457},
numpages = {9},
location = {Sydney NSW, Australia},
series = {WWW '25}
}

@misc{shao2024DeepSeekMathPushingLimits,
  title = {{{DeepSeekMath}}: {{Pushing}} the {{Limits}} of {{Mathematical Reasoning}} in {{Open Language Models}}},
  shorttitle = {{{DeepSeekMath}}},
  author = {Shao, Zhihong and Wang, Peiyi and Zhu, Qihao and Xu, Runxin and Song, Junxiao and Bi, Xiao and Zhang, Haowei and Zhang, Mingchuan and Li, Y. K. and Wu, Y. and Guo, Daya},
  year = {2024},
  number = {arXiv:2402.03300},
  eprint = {2402.03300},
  primaryclass = {cs},
  publisher = {arXiv},
  langid = {english}
}

@inproceedings{sun2023ChatGPTGoodSearch,
  title = {Is {{ChatGPT Good}} at {{Search}}? {{Investigating Large Language Models}} as {{Re-Ranking Agents}}},
  shorttitle = {Is {{ChatGPT Good}} at {{Search}}?},
  booktitle = {Proceedings of the 2023 {{Conference}} on {{Empirical Methods}} in {{Natural Language Processing}}},
  author = {Sun, Weiwei and Yan, Lingyong and et al},
  year = {2023},
  publisher = {Association for Computational Linguistics},
  langid = {english}
}

@misc{touvron2023LLaMAOpenEfficienta,
  title = {{{LLaMA}}: {{Open}} and {{Efficient Foundation Language Models}}},
  shorttitle = {{{LLaMA}}},
  author = {Touvron, Hugo and Lavril, Thibaut and et al},
  year = {2023},
  number = {arXiv:2302.13971},
  eprint = {2302.13971},
  primaryclass = {cs},
  publisher = {arXiv},
  langid = {english}
}

@inproceedings{wei2023ChainofThoughtPromptingElicitsa,
author = {Wei, Jason and Wang, Xuezhi and et al},
title = {Chain-of-thought prompting elicits reasoning in large language models},
year = {2022},
isbn = {9781713871088},
publisher = {Curran Associates Inc.},
address = {Red Hook, NY, USA},
booktitle = {Proceedings of the 36th International Conference on Neural Information Processing Systems},
articleno = {1800},
numpages = {14},
location = {New Orleans, LA, USA},
series = {NIPS '22}
}

@misc{wu2022PracticeImprovingSearch,
  title = {Some {{Practice}} for {{Improving}} the {{Search Results}} of {{E-commerce}}},
  author = {Wu, Fanyou and Liu, Yang and et al},
  year = {2022},
  number = {arXiv:2208.00108},
  eprint = {2208.00108},
  primaryclass = {cs},
  publisher = {arXiv},
  langid = {english}
}

@inproceedings{wu2025RankCoTRefiningKnowledge,
    title = "{R}ank{C}o{T}: Refining Knowledge for Retrieval-Augmented Generation through Ranking Chain-of-Thoughts",
    author = "Wu, Mingyan  and
      Liu, Zhenghao  and et al",
    editor = "Che, Wanxiang  and
      Nabende, Joyce  and
      Shutova, Ekaterina  and
      Pilehvar, Mohammad Taher",
    booktitle = "Proceedings of the 63rd Annual Meeting of the Association for Computational Linguistics (Volume 1: Long Papers)",
    month = jul,
    year = "2025",
    address = "Vienna, Austria",
    publisher = "Association for Computational Linguistics",
    url = "https://aclanthology.org/2025.acl-long.629/",
    doi = "10.18653/v1/2025.acl-long.629",
    pages = "12857--12874",
    ISBN = "979-8-89176-251-0",
}

@inproceedings{
xiong2021ANSWERINGCOMPLEXOPENDOMAIN,
title={Answering Complex Open-Domain Questions with Multi-Hop Dense Retrieval},
author={Wenhan Xiong and Xiang Li and et al},
booktitle={International Conference on Learning Representations},
year={2021},
url={https://openreview.net/forum?id=EMHoBG0avc1}
}

@inproceedings{
xiong2021APPROXIMATENEARESTNEIGHBOR,
title={Approximate Nearest Neighbor Negative Contrastive Learning for Dense Text Retrieval},
author={Lee Xiong and Chenyan Xiong and et al},
booktitle={International Conference on Learning Representations},
year={2021},
url={https://openreview.net/forum?id=zeFrfgyZln}
}

@inproceedings{yu2021FewShotConversationalDense,
  title = {Few-{{Shot Conversational Dense Retrieval}}},
  booktitle = {Proceedings of the 44th {{International ACM SIGIR Conference}} on {{Research}} and {{Development}} in {{Information Retrieval}}},
  author = {Yu, Shi and Liu, Zhenghao and et al},
  year = {2021},
  publisher = {ACM},
  copyright = {https://www.acm.org/publications/policies/copyright\_policy\#Background},
  langid = {english}
}

@inproceedings{yu2023FolkScopeIntentionKnowledgea,
  title = {{{FolkScope}}: {{Intention Knowledge Graph Construction}} for {{E-commerce Commonsense Discovery}}},
  shorttitle = {{{FolkScope}}},
  booktitle = {Findings of the {{Association}} for {{Computational Linguistics}}: {{ACL}} 2023},
  author = {Yu, Changlong and Wang, Weiqi and et al},
  year = {2023},
  publisher = {Association for Computational Linguistics},
  langid = {english}
}

@inproceedings{yu2024COSMOLargeScaleEcommercec,
  title = {{{COSMO}}: {{A Large-Scale E-commerce Common Sense Knowledge Generation}} and {{Serving System}} at {{Amazon}}},
  shorttitle = {{{COSMO}}},
  booktitle = {Companion of the 2024 {{International Conference}} on {{Management}} of {{Data}}},
  author = {Yu, Changlong and Liu, Xin and et al},
  year = {2024},
  publisher = {ACM},
  copyright = {https://www.acm.org/publications/policies/copyright\_policy\#Background},
  langid = {english}
}

@inproceedings{zhuang2024SetwiseApproachEffective,
  title = {A {{Setwise Approach}} for {{Effective}} and {{Highly Efficient Zero-shot Ranking}} with {{Large Language Models}}},
  booktitle = {Proceedings of the 47th {{International ACM SIGIR Conference}} on {{Research}} and {{Development}} in {{Information Retrieval}}},
  author = {Zhuang, Shengyao and Zhuang, Honglei and et al},
  year = {2024},
  publisher = {ACM},
  copyright = {https://creativecommons.org/licenses/by-nc-sa/4.0/},
  langid = {english}
}

@misc{zhuang2025RankR1EnhancingReasoning,
  title = {Rank-{{R1}}: {{Enhancing Reasoning}} in {{LLM-based Document Rerankers}} via {{Reinforcement Learning}}},
  shorttitle = {Rank-{{R1}}},
  author = {Zhuang, Shengyao and Ma, Xueguang and et al},
  year = {2025},
  number = {arXiv:2503.06034},
  eprint = {2503.06034},
  primaryclass = {cs},
  publisher = {arXiv},
  langid = {english}
}

@inproceedings{zou2022DivideConquerText,
  title = {Divide and {{Conquer}}: {{Text Semantic Matching}} with {{Disentangled Keywords}} and {{Intents}}},
  shorttitle = {Divide and {{Conquer}}},
  booktitle = {Findings of the {{Association}} for {{Computational Linguistics}}: {{ACL}} 2022},
  author = {Zou, Yicheng and Liu, Hongwei and et al},
  year = {2022},
  publisher = {Association for Computational Linguistics},
  langid = {english}
}

@InProceedings{Chen2022WANDSDatasetProduct,  
  title = {WANDS: Dataset for Product Search Relevance Assessment},  
  author = {Chen, Yan and Liu, Shujian and et al},  
  booktitle = {Proceedings of the 44th European Conference on Information Retrieval},  
  year = {2022},  
  numpages = {12}  
}

@inproceedings{wang2024ChainThoughtReasoning,
author = {Wang, Xuezhi and Zhou, Denny},
title = {Chain-of-thought reasoning without prompting},
year = {2025},
isbn = {9798331314385},
publisher = {Curran Associates Inc.},
address = {Red Hook, NY, USA},
booktitle = {Proceedings of the 38th International Conference on Neural Information Processing Systems},
articleno = {2123},
numpages = {27},
location = {Vancouver, BC, Canada},
series = {NIPS '24}
}

@article{zhao2023LargeLanguageModels,
  title={Large language models as commonsense knowledge for large-scale task planning},
  author={Zhao, Zirui and Lee, Wee Sun and et al},
  journal={Advances in neural information processing systems},
  volume={36},
  pages={31967--31987},
  year={2023}
}

@article{douze2025Faisslibrary,
      title={The Faiss library},
      author={Matthijs Douze and Alexandr Guzhva and et al},
      year={2024},
      eprint={2401.08281},
      archivePrefix={arXiv},
      primaryClass={cs.LG}
}

@inproceedings{ma2023FineTuningLLaMAMultiStagea,
author = {Ma, Xueguang and Wang, Liang and et al},
title = {Fine-Tuning LLaMA for Multi-Stage Text Retrieval},
year = {2024},
isbn = {9798400704314},
publisher = {Association for Computing Machinery},
address = {New York, NY, USA},
url = {https://doi.org/10.1145/3626772.3657951},
doi = {10.1145/3626772.3657951},
booktitle = {Proceedings of the 47th International ACM SIGIR Conference on Research and Development in Information Retrieval},
pages = {2421–2425},
numpages = {5},
location = {Washington DC, USA},
series = {SIGIR '24}
}

@misc{li2023GeneralTextEmbeddings,
  title = {Towards {{General Text Embeddings}} with {{Multi-stage Contrastive Learning}}},
  author = {Li, Zehan and Zhang, Xin and et al},
  year = {2023},
  number = {arXiv:2308.03281},
  eprint = {2308.03281},
  primaryclass = {cs},
  publisher = {arXiv},
  langid = {english}
}

@misc{he2021debertav3,
      title={DeBERTaV3: Improving DeBERTa using ELECTRA-Style Pre-Training with Gradient-Disentangled Embedding Sharing}, 
      author={Pengcheng He and Jianfeng Gao and et al},
      year={2021},
      eprint={2111.09543},
      archivePrefix={arXiv},
      primaryClass={cs.CL}
}

@misc{grattafiori2024llama3herdmodels,
      title={The Llama 3 Herd of Models}, 
      author={Aaron Grattafiori and Abhimanyu Dubey and et al},
      year={2024},
      eprint={2407.21783},
      archivePrefix={arXiv},
      primaryClass={cs.AI},
      url={https://arxiv.org/abs/2407.21783}, 
}

@inproceedings{ke2017lightgbm,
  title={LightGBM: A Highly Efficient Gradient Boosting Decision Tree},
  author={Ke, Guolin and Meng, Qi and et al},
  booktitle={Advances in Neural Information Processing Systems},
  volume={30},
  year={2017}
}

@inproceedings{zaheer2021bigbirdtransformerslonger,
author = {Zaheer, Manzil and Guruganesh, Guru and Dubey, Avinava and Ainslie, Joshua and Alberti, Chris and Ontanon, Santiago and Pham, Philip and Ravula, Anirudh and Wang, Qifan and Yang, Li and Ahmed, Amr},
title = {Big bird: transformers for longer sequences},
year = {2020},
isbn = {9781713829546},
publisher = {Curran Associates Inc.},
address = {Red Hook, NY, USA},
booktitle = {Proceedings of the 34th International Conference on Neural Information Processing Systems},
articleno = {1450},
numpages = {15},
location = {Vancouver, BC, Canada},
series = {NIPS '20}
}

@inproceedings{meng2021cocolmcorrectingcontrastingtext,
author = {Meng, Yu and Xiong, Chenyan and Baja, Payal and Tiwary, Saurabh and Bennett, Paul and Han, Jiawei and Song, Xia},
title = {COCO-LM: correcting and contrasting text sequences for language model pretraining},
year = {2021},
isbn = {9781713845393},
publisher = {Curran Associates Inc.},
address = {Red Hook, NY, USA},
booktitle = {Proceedings of the 35th International Conference on Neural Information Processing Systems},
articleno = {1769},
numpages = {13},
series = {NIPS '21}
}

\appendix
\section{Appendix}

\subsection{Statistical Analysis} \label{appx:Statistical Analysis}
\begin{figure}[h]
    \centering
    \includegraphics[width=1.01\columnwidth]{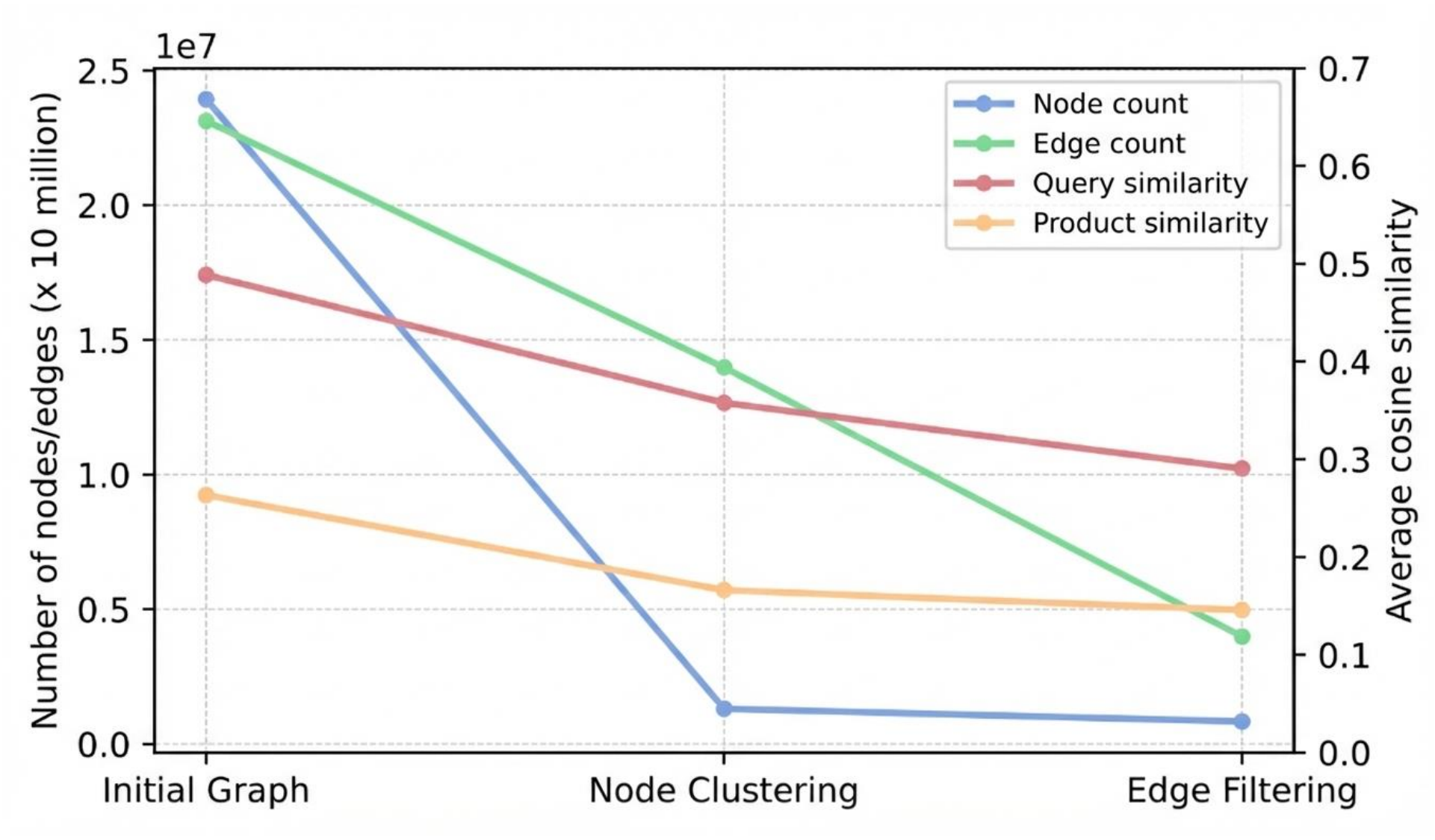}
    \caption{Statistics of the reasoning factor graph for the ESCI dataset along the E-CARE pipeline. Both the graph size and the in-group similarity of factors connected to each query or product node decrease as the pipeline progresses, indicating that node clustering and edge filtering reduce factor redundancy within groups, potentially facilitating downstream tasks.}
    \label{img:graph_stat}
\end{figure}

To evaluate the effectiveness of the E-CARE pipeline, we perform a statistical analysis examining how the structural and semantic properties of the reasoning factor graph evolve along the pipeline. Specifically, we analyze the number of nodes, the number of edges, and the in-group similarity of query and product nodes. The in-group similarity of a node on the reasoning factor graph is computed as 
\begin{equation}
    s = \frac{1}{|\sN|}\sum_{n \in \sN}{\frac{1}{|\sT|}\sum_{\sS\in\sT}{\frac{1}{|\sR_{\sS, n}|}\sum_{i,j \in \sR_{\sS. n},\, i \neq j}{\text{sim}(i,j)}}},
\end{equation}
where $\sN$ denotes the set of query or product nodes, $\sT$ represents the set of subsets of factors (defined in \S~\ref{sec:node_clustering}), and $\sR_{\sS,n}$ is the set of factors connected to node $n$ within $\sS\in\sT$ in $\mathcal{G}$. The results are presented in Figure \ref{img:graph_stat}. As shown, both the graph size and the in-group similarity of factors connected to each query or product node decrease progressively along the E-CARE pipeline, indicating that the factors on the final reasoning factor graph are less redundant and more distinguishable than those on the initial one. This low correlation among factors in the graph enriches the reasoning space and improves performance on downstream tasks.

\subsection{Adapter Results Evaluation} \label{Adapter Results Evaluation}

We compute the similarity score $s^k$ to measure how well the adapter’s top-$k$ outputs align with ground-truth positives across queries on the reasoning factor graph $\mathcal{G}$ as follows.

\begin{equation*}
    s^k = \frac{1}{|\sQ|}\sum_{q\in\sQ} \frac{1}{|a^k_{\sS}(q)|}\sum_{f\in a^k_{\sS}(q)}\text{max}\left[\text{sim}(f, n)\ |\ n \in P^+_{\sS}(q) \right]
\end{equation*}

For each query $q$, we take its top-$k$ results $a_{\sS}^k(q)$. Each result $f$ is matched to its most similar positive in $P^+_{\sS}(q)$, and these maxima are averaged over the $k$ results, followed by another averaging across all queries, yielding $s^k$.

\begin{figure}[t]
    \centering
\includegraphics[width=0.9\columnwidth]{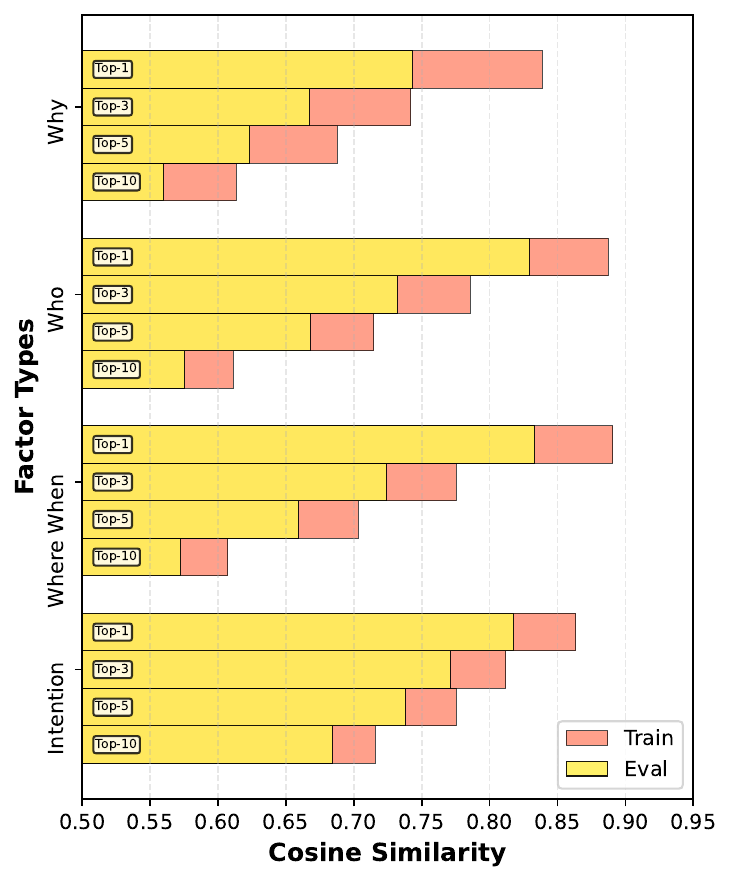}
    \caption{Similarity evaluation of product adapter results regarding 4 different factor subsets. The similarities are computed between text embeddings of predicted factors and ground truth factors.}
    \label{adapter_eval_product}
\end{figure}

The similarity evaluation results regarding the product adapter are shown in Figure \ref{adapter_eval_product}, which reaches the maximum cosine similarity of 0.89 in `where\_when' type at the top-1 setting on the training set.

\subsection{Inherent Reasoning Types and Extraction Example} \label{appx:Product_Features_Extraction}

An example of inherent LLM reasoning is shown in Table \ref{tab:extraction_example}.

The pre-defined product feature types and corresponding descriptions are shown in Table \ref{tab:product_features}.

\begin{table}[htbp]
    \centering
    \caption{An example of product feature extraction.}
    \begin{tabular}{|l|p{4cm}|}
    \hline
    \textbf{Product Name} & Panasonic FV-08VRE2 Ventilation Fan with Recessed LED (Renewed) \\ \hline
    \textbf{Extraction} & 
    category: Ventilation Fan; 
    
    style: Modern; 
    
    usage: Ventilation and Lighting \\ \hline
    \end{tabular}
    
    \label{tab:extraction_example}
\end{table}

\FloatBarrier

\begin{table}[htbp]
    \centering
    \caption{Pre-defined inherent reasoning types and corresponding descriptions.}
    \begin{tabular}{|l|p{4cm}|}
    \hline
    \textbf{Inherent Reasoning Type} & \textbf{Feature Type Description} \\ \hline
    category & The high-level functional taxonomy of the product, describing what kind of item it is and its primary role (e.g., running shoes, office chair, wireless earbuds). \\ \hline
    style & The aesthetic or design characteristics of the product, including visual appearance, material, and overall presentation (e.g., minimalist, vintage, sporty). \\ \hline
    usage & The intended usage scenarios or practical functions of the product, describing how and in what situations it is typically used (e.g., for long-distance running, for home office work, for outdoor travel). \\ \hline
\end{tabular}
    \label{tab:product_features}
\end{table}

\FloatBarrier

\subsection{LLM Commonsense Reasoning Prompt}
\label{appx:LLM Commonsense Reasoning Prompt}

Prompt template for scopt `Who' is shown in Table \ref{tab:prompt_template_who}. The `extraction\_response' is extracted by LLMs from the text description.


\begin{table}[htbp]
    \centering
    \small
    \caption{Prompt template for cross reasoning type `Who'. \{query\}, \{product\}, and \{extraction\_response\} are the placeholders for query text, product title text, and product extracted features text, respectively.}
    \begin{tabular}{|p{8cm}|}
    \hline
    \# Instruction:
    
    Given a <query> from user and the <product> that user clicked, your task is to answer the question in term of how user's <need> behind the <query> can be satisfied by <product>'s <utility>. 
    
    The <need> and <utility> within the answer should be less than 4 words. 
    
    Answer should be about type of person.
    
    Return 1 answer as least, 2 at maximum. 
    \newline
    
    \# Example 1:
    
    <query>: bachelorette vinyl stickers
    
    <product>: 
    
    title: Wedding Party Bridesmaid Vinyl Decal ONLY Set of 9 DIY Tumbler Cup Champagne Glasses Maid of Honor Gift (Metallic Gold) 
    
    category: Wedding Accessories
    
    broad\_category: Special Occasion Accessories
    
    target\_audience: Wedding Party
    
    shape: Rectangular
    
    size: 3.8" by 1.7"
    
    style: Gold Metallic
    
    quantity: 9
    
    material: Adhesive Vinyl
    
    usage: Hand wash only, removable but not reusable
    
    compatibility: Hard surface
    
    included\_accessories: Application Instructions
    
    Q: Given <query>, who will use <product>?
    
    A1: <product> will be used by [bridesmaid], which satisfies user's intention of [wedding decoration].
    
    A2: <product> will be used by [wedding planner], which 
    satisfies user's intention of buying [wedding preparation].
    \newline
        
    \ldots[More examples]\ldots
    \newline

    \# Example 4:
                
    <query>: \{query\}
    
    <product>: 
    
    title: \{product\}
    
    \{extraction\_response\}
    
    Q: Given <query>, who will use <product>? 
    \\ \hline
    \end{tabular}
    \label{tab:prompt_template_who}
\end{table}

\subsection{Edge Filtering Prompt} \label{Edge_Filtering_Prompt}

Some prompt templates for edge filtering are shown in Table \ref{tab:llm_self_evaluation_p2who}, \ref{tab:llm_self_evaluation_p2where_when}, \ref{tab:llm_self_evaluation_q2who}, and \ref{tab:llm_self_evaluation_q2where_when}, respectively. The remaining prompt templates follow the same format.

\begin{table}[h]
    \centering
    \small
    \caption{Prompt template of filtering product-to-who of edges, where `\{product\}' and `\{factor\}' are placeholders for `product' and `who' factor, respectively.}
    \begin{tabular}{|p{8cm}|}
    \hline
    
    \# Instruction:
    
    You are a labeling assistant, helping to clean invalid data. Please answer the following questions correctly. If correct, return YES, otherwise return NO.
    
    Just return YES or NO, don't return anything else.
    \newline
    
    \# examples:
    
    The product 'Jimmy Choo womens handbag white leather grained mini satchel' will be used by 'students'. Is this judgement reasonable? NO
    
    The product 'Rumikrafts Handmade Floral Trinket box heart shaped, Valentine gift for her' will be used by 'jewelley owner'. Is this judgement reasonable? YES
    
    \ldots[More examples]\ldots
    \newline

    The product '\{product\}' will be used by '\{factor\}'. Is this judgment reasonable?
    \\ \hline
    \end{tabular}
    \label{tab:llm_self_evaluation_p2who}
\end{table}

\begin{table}[h]
    \centering
    \small
    \caption{Prompt template of filtering product-to-where\_when of edges, where `\{product\}' and `\{factor\}' are placeholders for `product' and `where\_when' factor, respectively.}
    \begin{tabular}{|p{8cm}|}
    \hline
    
    \# Instruction:
    
    You are a labeling assistant, helping to clean invalid data. Please answer the following questions correctly. If correct, return YES, otherwise return NO.
    
    Just return YES or NO, don't return anything else.
    \newline
    
    \# examples:
    
    The product 'French A1 to B2: A complete guide' will be used in the 'language learning' scenario. Is this reasonable? YES
    
    The product 'Arcteryx snow sports cargo pants XX\_Large 32' will be used in the 'hiking' scenario. Is this reasonable? NO
    
    \ldots[More examples]\ldots
    \newline

    The product '\{product\}' will be used in the '\{factor\}' scenario. Is this reasonable?
    \\ \hline
    \end{tabular}
    \label{tab:llm_self_evaluation_p2where_when}
\end{table}

\begin{table}[h]
    \centering
    \small
    \caption{Prompt template of filtering query-to-who of edges, where `\{query\}' and `\{factor\}' are placeholders for `query' and `who' factor, respectively.}
    \begin{tabular}{|p{8cm}|}
    \hline
    
    \# Instruction:
    
    You are a labeling assistant, helping to clean invalid data. Please answer the following questions correctly. If correct, return YES, otherwise return NO.
    
    Just return YES or NO, don't return anything else.
    \newline
    
    \# examples:
    
    The user searched for 'Electronic drum set for kids', which means the user is a 'beginner'. Is this reasonable? YES
    
    The user searched for 'Arcteryx snow sports cargo pants', which means the user is a 'beach lover'. Is this reasonable? NO

    \ldots[More examples]\ldots
    \newline

    The user searched for '\{query\}', which means the user is a '\{factor\}'. Is this reasonable?
    
    \\ \hline
    \end{tabular}
    \label{tab:llm_self_evaluation_q2who}
\end{table}

\begin{table}[h]
    \centering
    \small
    \caption{Prompt template of filtering query-to-where\_when of edges, where `\{query\}' and `\{factor\}' are placeholders for `query' and `where\_when' factor, respectively.}
    \begin{tabular}{|p{8cm}|}
    \hline
    
    \# Instruction:
    
    You are a labeling assistant, helping to clean invalid data. Please answer the following questions correctly. If correct, return YES, otherwise return NO.
    
    Just return YES or NO, don't return anything else.
    \newline
    
    \# examples:
    
    The user searched for '\#2 pencils HB wood cased', indicating that the user's usage scenario is 'going out'. Is this reasonable? NO
    
    The user searched for '\#2 pencils HB wood cased', indicating that the user's usage scenario is 'classroom'. Is this reasonable? YES

    \ldots[More examples]\ldots
    \newline

    The user searched for '\{query\}', which indicates that the user's usage scenario is '\{factor\}'. Is this reasonable?
    
    \\ \hline
    \end{tabular}
    \label{tab:llm_self_evaluation_q2where_when}
\end{table}

\subsection{Noe Clustering Prompt} \label{Prompt_of_Edge_Clustering}

Table \ref{tab:prompt_for_clustering} shows the prompt we use for factors clustering with LLM.

\begin{table}[h]
    \centering
    \small
    \caption{The prompt template of clustering, where the \{factors\} is the placeholder for the input factor list.}
    \begin{tabular}{|p{8cm}|}
    \hline
    
    \# Instruction:
    1. Use a summary phrase to summarize the provided phrase list.
    2. The general phrase should be less than 2 words. 
    3. Only the general phrase part is output, but the phrase list part is not output.
    \newline
    
    \# Example 1:
    
    - Phrase list: [slip on shoes, loafer shoes]
    
    - General phrase: slip-on loafer
    
    \# Example 2:
    
    - Phrase List: [cotton t-shirt, breathable t-shirt, t-shirt made of cotton]
    
    - General phrase: breathable cotton t-shirt

    \ldots[More examples]\ldots
    \newline
    
    \# Example 6:
    
    - Phrase List: \{factors\}
    
    - General phrase:
    \\ \hline
    \end{tabular}
    \label{tab:prompt_for_clustering}
\end{table}

\FloatBarrier

\end{document}